\documentclass[11pt]{article}

\usepackage[margin=1in]{geometry}

\usepackage{graphics}
\usepackage{epsfig}
\usepackage{epstopdf}
\usepackage{amsmath}
\usepackage{array}
\begin{document}
\begin{center}
{\LARGE Phase space analysis of bulk viscous matter dominated
universe \\[0.2in]}
{ Athira Sasidharan and Titus K Mathew\\
e-mail:athirasnair91@cusat.ac.in, titus@cusat.ac.in \\ Department of
Physics, Cochin University of Science and Technology, Kochi, India.}
\end{center}

\date{\today}

\abstract{We consider a Friedmann model of the universe with bulk
viscous matter and radiation as the cosmic components. We study the
asymptotic properties in the equivalent phase space by considering
the three cases for the bulk viscous coefficient as (i)
$\zeta=\zeta_{0}$, a constant (ii)
$\zeta=\zeta_{0}+\zeta_{1}\frac{\dot{a}}{a}$, depending on velocity
of the expansion of the universe and (iii)
$\zeta=\zeta_{0}+\zeta_{1}\frac{\dot{a}}{a}+\zeta_{2}\frac{\ddot{a}}{\dot{a}}$,
depending both on velocity and acceleration of the expansion of the
universe. It is found that all the three cases predicts the late
acceleration of the universe. However, a conventional realistic
behaviour of the universe, i.e., a universe having an initial
radiation dominated phase, followed by decelerated matter dominated
phase and then finally evolving to accelerated epoch, is shown only
when $\zeta=\zeta_{0}$, a constant. For the other two cases, it does
not show either a prior conventional radiation dominated phase or a
matter dominated phase of the universe.}

\section{Introduction}
\label{intro} From the observations on Type I a supernova
~\cite{Riess1,Perl1}, it is clear that the present universe is
undergoing an accelerated expansion. This was further confirmed by
the observations on cosmic microwave background radiations (CMBR)
~\cite{Bennet1}, large scale structure (LSS) ~\cite{Tegmark1}, the
Sloan Digital Sky Survey (SDSS) ~\cite{Seljak}, the Wilkinson
Microwave Anisotropy Probe (WMAP) ~\cite{Komatsu1}, etc. Many models
have been introduced to explain the current acceleration. Basically
there are two approaches - one is to propose suitable forms for the
energy-momentum tensor $T_{\mu \nu}$ in the Einstein's equation,
having a negative pressure, which culminate in the proposal of an
exotic energy called dark energy. The second approach is to modify
the geometry of the space time in the Einstein's equation. Among the
models of dark energy, the simplest candidate is the cosmological
constant ~\cite{Weinberg1}. However, it suffers from the coincidence
problem and the fine tuning problem ~\cite{Copeland1}. So as a
result, dynamical dark energy models such as quintessence
~\cite{fujii,carroll,ford,Liddle}, k-essence
~\cite{chiba1,Picon2000} and perfect fluid models (like Chaplygin
gas model) ~\cite{kamen1,Bento2002} were considered. By modifying
the geometry of space time, models such as $f(R)$ gravity
~\cite{capo1,Thomas2010}, $f(T)$\ gravity
~\cite{ferraro1,Ratbay2011}, Gauss-Bonnet theory ~\cite{nojiri},
Lovelock gravity ~\cite{pad2}, Horava-Lifshitz gravity
~\cite{horava1}, scalar-tensor theories ~\cite{amendola1},
braneworld models ~\cite{dvali1} etc, have been proposed.

In the context of inflation, many authors found that the bulk
viscous fluids are capable of producing acceleration of the universe
~\cite{pad,waga,Cheng}. This idea was extended to explain the late
acceleration of the universe ~\cite{fabris1,li1,Hiplito1,av1,av2}.
Of the fluid dissipative phenomena, bulk viscosity is the most
favorable phenomenon, compatible with the symmetry requirements of
the homogeneous and isotropic Friedmann-Lemaitre-Robertson-Walker
(FLRW) universe. The bulk viscosity can be considered as a measure
of the pressure required to restore equilibrium when the cosmic
fluid expands in an expanding universe. In ~\cite{wilson1}, the
authors have considered a mechanism for the formation of bulk
viscosity by the decay of a dark matter particle into relativistic
products.

In an earlier work ~\cite{athira}, we have analyzed the cosmic
evolution of the bulk viscous matter dominated universe with bulk
viscous coefficient $\zeta$ depending on both the velocity and
acceleration of the expanding universe as,
$\zeta=\zeta_{0}+\zeta_{1}\frac{\dot{a}}{a}+\zeta_{2}\frac{\ddot{a}}{\dot{a}}$,
where $a$ is the scale factor of expansion of the universe. The
model predicts the late acceleration of the universe with transition
redshift around $z_T\sim0.49$. The model also predicts the present
deceleration parameter around $-0.68$, which is very much in
agreement with the observational result, around $-0.64$
~\cite{WMAP2009}. The present paper concentrate on the phase space
analysis of the model. A phase space analysis of a cosmological
model would indicate the different stages of the universe like a) a
radiation dominated phase, followed by b) a matter dominated phase,
and c) an accelerated expanding phase, corresponding to the
existence of different critical points. So doing a phase space
analysis would clearly indicate whether the model predicts the
realistic picture regarding the evolution of our universe.

The paper is organized as follows: In Section ~\ref{sec:1}, we
present the basic formalism of the bulk viscous universe. In Section
~\ref{sec:2}, we consider a flat universe containing bulk viscous
matter alone (neglecting radiation) and we estimate the values of
the bulk viscous parameters corresponding to each case by
contrasting the model with the Type I a supernova data and presented
the phase space analysis of the model. In Section ~\ref{sec:3}, we
discuss a flat universe containing both the radiation and bulk
viscous matter as the cosmic components and did a phase space
analysis of the model for each cases separately. Finally, we present
our conclusion in Section ~\ref{sec:4}.

\section{Bulk viscous universe}
\label{sec:1} We consider a spatially flat universe described by the
Friedmann-Lemaitre-Robertson-Walker (FLRW) metric,
\begin{equation}
ds^{2}=-dt^{2}+a(t)^{2}(dr^{2}+r^{2}d\theta ^{2}+r^{2}\sin
^{2}\theta d\phi ^{2})
\end{equation}
where $t$ is cosmic time, $a(t)$ is the scale factor and
$(r,\theta,\phi)$ are the comoving spacial co-ordinates. Using the
Eckart formalism ~\cite{Eckart1}, the effective pressure of the bulk
viscous fluid is
\begin{equation}
\label{p} P^{*}=P-3\zeta H
\end{equation}
where $P$ is the normal kinetic pressure and $\zeta$ is the
coefficient of bulk viscosity. Let us assume that bulk viscous fluid
is the non-relativistic matter with $P=0$ and so the contribution to
effective pressure is only due to the negative viscous pressure. A
more general theory for bulk viscous stress was proposed by Israel
and Stewart ~\cite{Israel2,Israel3} and one could obtain the Eckart
theory from it, in the limit of vanishing relaxation time. So, in
this limit, the Eckart theory is a good approximation to the
Israel-Stewart theory. Eckart's theory consider only the first order
deviation from equilibrium and neglects the second order terms,
while the theory developed by Israel and Stewart is a second order
theory. Even though Eckart theory suffers from causality problems,
it is the simplest alternative and is less complicated than the
Israel-Stewart theory. So it has been used widely by many authors to
characterize the bulk viscous fluid. For example in Refs.
~\cite{fabris1,kremer2003,titus,Hu2006,Gron2013,Brevik2005,Colistete2007},
the Eckart approach has been used in dealing with the accelerating
universe with the bulk viscous fluid. We followed the Eckart
formalism for the viscous pressure.

The Friedmann equations describing the evolution of a flat universe
dominated with bulk viscous matter are
\begin{equation}
\label{friedmann}\left(\frac{\dot{a}}{a}\right)^2=\frac{\rho}{3}
\end{equation}
\begin{equation}
\label{friedmanna}
2\frac{\ddot{a}}{a}+\left(\frac{\dot{a}}{a}\right)^{2}=-P^{*}
\end{equation}
where we have taken $8\pi G = 1$, $\rho$ is the density of the
content of the universe and overdot represents the derivative with
respect to cosmic time $t$. The conservation equation is
\begin{equation}
\label{conser} \dot{\rho}+3H\left(\rho+P^{*}\right)=0.
\end{equation}

From the Fluid mechanics, it is clear that the bulk viscosity
coefficient, $\zeta$ is related to the rate of compression or
expansion of the fluid \cite{Lie}. In the present model, the fluid
is comoving with the expanding universe. So, the velocity and
acceleration of the fluid is the same as that of the expanding
universe, which are $\dot{a}$ and $\ddot{a}$, respectively. Since
there is no conclusive microscopic theory to calculate the transport
coefficient, it is logical to consider $\zeta$ to be depending on
the velocity and acceleration, $\dot{a}$ and $\ddot{a}$. The best
way is to take a linear combination of the three terms: the first
term a constant $\zeta_{0}$, the second proportional to the velocity
and the third proportional to the acceleration
\cite{ren1,Singh,Avelino,athira} as,
\begin{equation}
\label{zeta}
\zeta=\zeta_{0}+\zeta_{1}\frac{\dot{a}}{a}+\zeta_{2}\frac{\ddot{a}}{\dot{a}}=\zeta_{0}+\zeta_{1}H+\zeta_{2}(\frac{\dot{H}}{H}+H).
\end{equation}
On taking this form of time dependent bulk viscosity, the equation
of state assumes the most general form
\cite{ren1,Nojiri,Nojiri1,Xu},
\begin{equation}
\label{general} P_{eff}=\omega
\rho+P_{0}+w_{H}H+w_{H2}H^{2}+w_{dH}\dot{H}
\end{equation}
By comparing Eqs.(\ref{general}), (\ref{p}) and (\ref{zeta}), we
could identify, $w_{H}=-3\zeta_{0}$,
$w_{H2}=-3(\zeta_{1}+\zeta_{2})$ and $w_{dH}=-3\zeta_{2}$.

In this paper, we consider in detail the following three cases.

\begin{enumerate}
\item[Case 1] with $\tilde{\zeta}_0$, $\tilde{\zeta}_1$ and
$\tilde{\zeta}_2$ all nonzero, so that the total bulk viscous
parameter
$\zeta=\zeta_{0}+\zeta_{1}\frac{\dot{a}}{a}+\zeta_{2}\frac{\ddot{a}}{\dot{a}}$,
depending on both the velocity and acceleration of the expansion of
the universe.
\item[Case 2] with $\tilde{\zeta}_2=0$, so
$\zeta=\zeta_{0}+\zeta_{1}\frac{\dot{a}}{a}$, depending only on
velocity of the expansion of the universe and not on its
acceleration
\item[Case 3] with
$\tilde{\zeta}_1=\tilde{\zeta}_2=0$, so $\zeta=\zeta_{0}$, a
constant
\end{enumerate}

where we define the dimensionless bulk viscous parameters
$\tilde{\zeta}_{0}$, $\tilde{\zeta}_{1}$ and $\tilde{\zeta}_{2}$ and
the total dimensionless viscous parameter $\tilde{\zeta}$ as,
\begin{equation}
\label{dimen} \tilde{\zeta}_{0}=\frac{3\zeta_{0}}{H_{0}},\ \ \ \
\tilde{\zeta}_{1}=3\zeta_{1},\ \ \ \ \tilde{\zeta}_{2}=3\zeta_{2},\
\ \ \ \tilde{\zeta}=\frac{3\zeta}{H_{0}}
\end{equation}

\section{Flat universe with bulk viscous matter}
\label{sec:2} In this section, we consider a universe dominated with
matter (neglecting radiation). The Friedmann equations
\eqref{friedmann} and \eqref{friedmanna} (by substituting
$H=\frac{\dot{a}}{a}$) becomes,
\begin{equation}
 H^{2}=\frac{\rho_{m}}{3}
\end{equation}
\begin{equation}
\label{dH1} 2\dot{H}+3H^{2}=H H_{0}\tilde{\zeta}
\end{equation}
and the conservation equation becomes,
\begin{equation}
\label{conser1} \dot{\rho_{m}}+3H\left(\rho_{m}-H
H_{0}\tilde{\zeta}\right)=0.
\end{equation}
where $H_{0}$ is the present value of the Hubble parameter and
$\rho_m$ is the matter density. Using the expression for $\dot{H}$
from Eq. \eqref{dH1}, we get the total dimensionless bulk viscous
parameter $\tilde{\zeta}$ (Eq. \eqref{zeta}) as,
\begin{equation}
\label{zeta1}
\tilde{\zeta}=\frac{1}{2-\tilde{\zeta}_{2}}\left[2\tilde{\zeta}_{0}+\left(2\tilde{\zeta}_{1}-\tilde{\zeta}_{2}\right)
\frac{H}{H_{0}}\right],
\end{equation}

The deceleration parameter $q$ and the equation of state parameter
$\omega$ are defined as,
\begin{equation}
\label{deceleration} q=-1-\frac{\dot{H}}{H^{2}}
\end{equation}
\begin{equation}
\label{equation} \omega=-1-\frac{2}{3}\frac{\dot{H}}{H^2}
\end{equation}

Using Eqs. \eqref{dH1} and \eqref{zeta1}, $q$ and $\omega$ becomes
\begin{equation}
\label{decelerationa}
q=\frac{1}{2-\tilde{\zeta}_{2}}\left(1-\tilde{\zeta}_{1}-\tilde{\zeta}_{0}\frac{H_0}{H}\right)
\end{equation}
\begin{equation}
\label{equationa}
\omega=\frac{1}{3(2-\tilde{\zeta}_2)}\left(\tilde{\zeta}_2-2\tilde{\zeta}_1-2\tilde{\zeta}_0\frac{H_0}{H}\right)
\end{equation}
The evolution of these parameters was studied and its present values
was extracted in our earlier work ~\cite{athira}.

The evolution of Hubble parameter can be obtained from Eqs.
\eqref{dH1} and \eqref{zeta} by replacing $dt$ with $\ln a$ and then
on integrating as,
\begin{equation}
\label{hubbleina}
H(a)=H_{0}\left[a^{\frac{\tilde{\zeta}_{1}+\tilde{\zeta}_{2}-3}{2-\tilde{\zeta}_{2}}}\left(1+\frac{\tilde{\zeta}_{0}}{\tilde{\zeta}_{1}+\tilde{\zeta}_{2}-3}\right)-\frac{\tilde{\zeta}_{0}}{\tilde{\zeta}_{1}+\tilde{\zeta}_{2}-3}\right]
\end{equation}
When $\tilde{\zeta}_0=\tilde{\zeta}_1=\tilde{\zeta}_2=0$, $H$
reduces to $H_0a^{-\frac{3}{2}}$, which corresponds to the ordinary
matter dominated universe. The Hubble parameter governs the
behaviour of deceleration parameter and equation of state parameter.

\subsection{Parameter estimation using Type Ia Supernova}
\label{sec:par} In this section we estimate the values of
$\tilde{\zeta}_0$, $\tilde{\zeta}_1$ and $\tilde{\zeta}_2$ using SCP
``Union"  Type Ia Supernova data ~\cite{kowalski1} composed of 307
type Ia Supernovae from 13 independent data sets. In our earlier
work ~\cite{athira}, we have extracted the values of
$\tilde{\zeta}_0$, $\tilde{\zeta}_1$ and $\tilde{\zeta}_2$
simultaneously. Here, in addition to that we are evaluating the
coefficients as per the conditions mentioned in case 2 and case 3
respectively.

In a flat universe, the luminosity distance $d_{L}$ is defined as
\begin{equation}
d_{L}=c(1+z)\int_{0}^{z}\frac{dz'}{H}
\end{equation}
where $H$ is the Hubble parameter, $c$ is the speed of light and $z$
is the redshift. The theoretical distance moduli $\mu_{t}$ for the
k-th Supernova with redshift $z_{k}$ is given as,
\begin{equation}
\mu_{t}=m-M=5\log_{10}[\frac{d_{L}}{Mpc}]+25
\end{equation}
where, $m$ and $M$ are the apparent and absolute magnitudes of the
SNe respectively. Then we can construct $\chi^{2}$ function as,
\begin{equation}
\chi^{2}\equiv
\sum^{n}_{k=1}\frac{\left[\mu_{t}-\mu_{k}\right]^{2}}{\sigma_{k}^{2}}
\end{equation}
where $\mu_{k}$ is the observational distance moduli for the k-th
Supernova, $\sigma_{k}^{2}$ is the variance of the measurement and
$n$ is the total number of data, here $n=307$. The $\chi^{2}$
function, thus obtained is then minimized to obtain the best
estimate of the parameters, $\tilde{\zeta}_{0}$,
$\tilde{\zeta}_{1}$, $\tilde{\zeta}_{2}$ and $H_0$. For the first
case i.e., with
$\zeta=\zeta_{0}+\zeta_{1}\frac{\dot{a}}{a}+\zeta_{2}\frac{\ddot{a}}{\dot{a}}$,
we have evaluated the values of $\tilde{\zeta}_0$, $\tilde{\zeta}_1$
and $\tilde{\zeta}_2$ simultaneously in reference ~\cite{athira}.
For Case 1, we have considered both $\tilde{\zeta}_0<0$ and
$\tilde{\zeta}_0>0$, however both these cases leads to identical
evolution for the total $\tilde{\zeta}$. For details refer
~\cite{athira}. In addition to this, we evaluate the values of the
bulk viscous parameters corresponding to the cases - case 2 and case
3. These values are given in the table ~\ref{tab:1}. In order to
compare the results of the present model, we have also estimated the
values for $\Lambda$CDM model using the same data set. We find that
the values of $H_0$ and goodness-of-fit $\chi^2_{d.o.f.}$ for the
$\Lambda$CDM model are very close to those obtained from the present
bulk viscous model.

\begin{table}[tbp]
\centering
\begin{tabular}{|c|cc|c|c|c|}
\hline Cases & Case 1&& Case 2 & Case 3 & $\Lambda$CDM
\\ \hline
Conditions & $\tilde{\zeta}_0>0$ & $\tilde{\zeta}_0<0$ & $\tilde{\zeta}_0>0$ & $\tilde{\zeta}_0>0$ & - \\
\hline
$\tilde{\zeta}_0$ & 7.83 & -4.68 & 6.26 & 1.92 & - \\
$\tilde{\zeta}_1$ & -5.13 & 4.67 & -3.91 & 0 & - \\
$\tilde{\zeta}_2$ & -0.51 & 3.49 & 0 & 0 & - \\
$\Omega_{m0}$ & 1 & 1 & 1 & 1 & 0.316 \\
$H_0$ & 70.49 & 70.49 & 70.49 & 69.61 & 70.03 \\
$\chi^2_{min}$ & 310.54 & 310.54 & 310.54 & 315.07 & 311.93\\
$\chi^2_{d.o.f}$ & 1.02 & 1.01 & 1.02 & 1.03 & 1.02 \\
\hline
\end{tabular}
\caption{\label{tab:1} Best estimates of the Bulk viscous parameters
and $H_{0}$ and also $\chi^{2}$ minimum value corresponding to the
above different cases of $\zeta$.
$\chi^{2}_{d.o.f}=\frac{\chi^{2}_{min}}{n-m}$, where $n=307$, the
number of data and $m$ is the number of parameters in the model. The
subscript d.o.f stands for degrees of freedom. For the best
estimation we have use SCP ``Union" 307 SNe Ia data sets. The values
of parameter corresponding to the first case is extracted in
~\cite{athira}. We have also shown the best estimates for the
$\Lambda$CDM model for comparison, where $\Omega_{m0}$ is the
present mass density parameter.}
\end{table}

\subsection{Phase space analysis of bulk viscous matter dominated
universe} It is difficult to solve exactly the cosmological field
equations with more than one cosmic components. Often one make use
of the dynamical system tools to extract the asymptotic properties
of the model. For this we write down the cosmological equations as a
system of autonomous differential equations and then investigate the
equivalent phase space of the model. The critical points of these
equations can be correlated with the solutions of the cosmological
field equations and its stability can be determined by examining the
system obtained by linearizing about the critical point i.e., from
the eigen values of the corresponding Jacobian matrix. The first
step is to select suitable dynamic variables for the phase space
analysis. We consider $u$ and $v$ as the dimensionless phase space
variables which are defined as follows,
\begin{equation}
u=\Omega_{m}=\frac{\rho_{m}}{3H^{2}}
\end{equation}
\begin{equation}
v=\frac{1}{\frac{H_{0}}{H}+1}
\end{equation}
These phase space coordinates are varying in the range $0\leq
u\leq1$ and $0\leq v\leq1$.

\subsubsection{Case 1: with
$\zeta=\zeta_{0}+\zeta_{1}\frac{\dot{a}}{a}+\zeta_{2}\frac{\ddot{a}}{\dot{a}}$}
Using Friedmann equations, conservation equation for matter and Eq.
\eqref{zeta1}, we can obtain the autonomous equations satisfied by
$u$ and $v$ as
\begin{equation}
\label{eq:aut1}
u'=\frac{\left(1-u\right)\left(2\tilde{\zeta}_{0}\left(1-v\right)+\left(2\tilde{\zeta}_{1}-\tilde{\zeta}_{2}\right)v\right)}{v\left(2-\tilde{\zeta}_{2}\right)}=f(u,v)
\end{equation}
\begin{equation}
\label{eq:aut2}
v'=\frac{\left(1-v\right)\left(\tilde{\zeta}_{0}\left(1-v\right)+\left(\tilde{\zeta}_{1}+\tilde{\zeta}_{2}-3\right)v\right)}{2-\tilde{\zeta}_{2}}=g(u,v)
\end{equation}
where the prime denote the derivative with respect to $\ln a$. Using
Eqs. \eqref{decelerationa} and \eqref{equationa}, the deceleration
parameter and equation of state parameter can be written in terms of
$v$ as
\begin{equation}
\label{deceleration1}
q=\frac{1}{2-\tilde{\zeta}_{2}}\left(1-\tilde{\zeta}_{1}-\tilde{\zeta}_{0}\frac{1-v}{v}\right)
\end{equation}
\begin{equation}
\label{equation1}
\omega=\frac{1}{3(2-\tilde{\zeta}_2)}\left(\tilde{\zeta}_2-2\tilde{\zeta}_1-2\tilde{\zeta}_0\frac{1-v}{v}\right)
\end{equation}

The critical points $(u_c,v_c)$ of the above autonomous equations
Eqs. \eqref{eq:aut1} and \eqref{eq:aut2} can be obtained by equating
$u'=0$ and $v'=0$. The stability of the dynamic system in the
neighbourhood of the critical point can be checked as follows.
Linearize the system by considering small perturbations around the
critical point $u\rightarrow u_c+\delta u$, $v\rightarrow v_c+\delta
v$, which satisfy the following matrix equation,
\begin{equation}
\label{eq:matrixa}
\begin{bmatrix}
\delta u' \\ \delta v'
\end{bmatrix}
=\begin{bmatrix}\left(\frac{\partial f}{\partial u}\right)_0 &
 \left(\frac{\partial f}{\partial v}\right)_0
\\
\left(\frac{\partial g}{\partial u}\right)_0 & \left(\frac{\partial
g}{\partial v}\right)_0
\end{bmatrix}
\begin{bmatrix}
\delta u\\ \delta v
\end{bmatrix}
\end{equation}
where the suffix 0 denotes the value evaluated at the critical point
$(u_c,v_c)$. The Jacobian matrix ($2\times 2$ matrix in the right
hand side of the Eq. \eqref{eq:matrixa}) for the autonomous
equations Eq. \eqref{eq:aut1} and Eq. \eqref{eq:aut2} is
\begin{equation}
\label{eq:matrix}
\begin{bmatrix}
\left(-\frac{2\tilde{\zeta}_0(v-1)+(\tilde{\zeta}_2-2\tilde{\zeta}_1)v}{(\tilde{\zeta}_2-2)v}\right)_0
& \left(\frac{2\tilde{\zeta}_0(u-1)}{(\tilde{\zeta}_2-2)v^2}\right)_0 \\
0 &
\left(\frac{-2\tilde{\zeta}_0(v-1)+(\tilde{\zeta}_1+\tilde{\zeta}_2-3)(2v-1)}{\tilde{\zeta}_2-2}\right)_0
\end{bmatrix}\end{equation}
If the eigen values of the Jacobian matrix are all negative, then
the critical point is stable otherwise the critical point is
generally unstable. If all the eigen values are positive then the
critical point is an unstable node and if there are both positive
and negative eigen values, then the critical point is a saddle
point.

\begin{figure}[tbp]
\centering
\includegraphics[width=.45\textwidth]{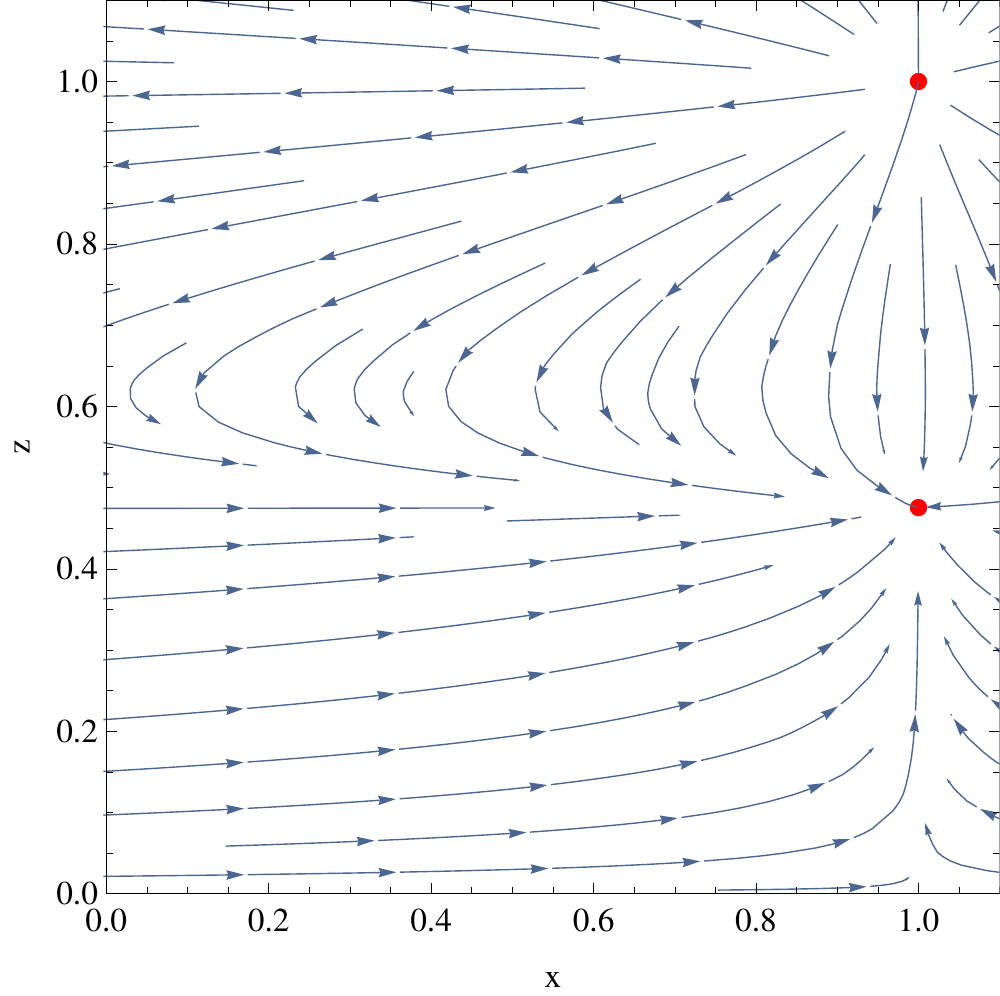}
\caption{\label{fig:1} The figure shows the phase space structure in
the $u-v$ plane corresponding to the Case 1
($\zeta=\zeta_{0}+\zeta_{1}\frac{\dot{a}}{a}+\zeta_{2}\frac{\ddot{a}}{\dot{a}}$).
The critical point (1,1) in the upper right corner of the plot is a
past attractor and the point (1,0.475), below the first critical
point, is a future attractor. The direction of the trajectories is
shown by the arrow head.}
\end{figure}

For autonomous equations \eqref{eq:aut1} and \eqref{eq:aut2}, there
are two critical points $(u_c,v_c):$
\begin{enumerate}
\item $(u_c,v_c)=(1,1)$ \\
Here $u=1$ implies a viscous matter dominated universe and $v=1$
corresponds either to $H_0=0$ or $H\rightarrow\infty$. Since $H_0$
cannot be zero, this corresponds to the initial singular state
characterized with $H\rightarrow\infty$. The Jacobian matrix
corresponding to this critical point can be obtained by putting
$u=1$ and $v=1$ in Eq. \eqref{eq:matrix}. The eigen values of the
Jacobian matrix are
\begin{equation}
\lambda_1=\frac{2\tilde{\zeta}_1-\tilde{\zeta}_2}{\tilde{\zeta}_2-2},\
\ \ \ \ \ \ \ \
\lambda_2=\frac{\tilde{\zeta}_1+\tilde{\zeta}_2-3}{\tilde{\zeta}_2-2}
\end{equation}
Substituting the values of $\tilde{\zeta}_0$, $\tilde{\zeta}_1$ and
$\tilde{\zeta}_2$ from Table ~\ref{tab:1}, we get $\lambda_1=3.88$
and $\lambda_2=3.44$ for the first condition (i,e.,
$\tilde{\zeta}_0>0$) and $\lambda_1=3.915$ and $\lambda_2=3.457$ for
the second condition (i,e., $\tilde{\zeta}_0<0$), which are almost
the same except for the slight difference in the decimal places.
Since both the eigen values are positive, the critical point is
unstable and is a past attractor. The values of equation of state
parameter $\omega$ and deceleration parameter $q$ (using Eqs.
\eqref{equation1} and \eqref{deceleration1}) are found to be around
$1.3$ and $2.4$ respectively, for the two cases. This shows that in
the early stage of the evolution of the universe, bulk viscous
matter will behave almost like a stiff fluid.
\item
$(u_c,v_c)=(1,\frac{\tilde{\zeta}_0}{\tilde{\zeta}_0-(\tilde{\zeta}_1+\tilde{\zeta}_2-3)})=(1,0.475)$
\\ This also corresponds to a matter dominated universe with
$\frac{H_0}{H}=1.105$. The eigen values corresponding to this point
are
\begin{equation}
\lambda_1=-\frac{\tilde{\zeta}_1+\tilde{\zeta}_2-3}{\tilde{\zeta}_2-2},\
\ \ \ \ \ \ \lambda_2=-3
\end{equation}
Using the values of bulk viscous parameters from Table ~\ref{tab:1},
we obtain $\lambda_1\sim-3.45$ for the two conditions (i.e., for
$\tilde{\zeta}_0<0$ and $\tilde{\zeta}_0>0$). Since the two eigen
values are negative, this critical point is a stable node and a
future attractor. It is found that $\omega\sim-1$ and $q\sim-1$ and
it corresponds to de-Sitter phase.
\end{enumerate}
The phase space plot for this case is shown in the Fig.
~\ref{fig:1}. From the figure, it is clear that the critical point
(1,1) is an unstable past attractor as trajectories emerge from this
point. These emerging trajectories finally converges to the critical
point (1,0.475), which is the future attractor. So the phase plot
analysis of this case suggest a universe which begins from an
initial singular state and ends on a de-Sitter type universe. This
is almost similar to the picture given by the $\Lambda$CDM model, in
which the universe evolves from an initial singularity to a de
Sitter phase through a matter dominated epoch. The critical points,
their stability and the values of $\omega$ and $q$ are summarized in
the Table ~\ref{tab:2}.

\begin{table}[tbp]
\centering
\begin{tabular}{|c|l|c|p{1.7cm}|l|l|}
\hline
$(u_c,v_c)$ & $\lambda_1$ & $\lambda_2$ & Stability & $\omega$ & $q$  \\
\hline
$(1,1)$ & 3.9 & 3.4 & Unstable, Past attractor & 1.3 & 2.4 \\
$(1,0.475)$ & -3.45 & -3 & Stable, future attractor & -1 & -1 \\
\hline
\end{tabular}
\caption{\label{tab:2} Critical points for case 1, with
$\zeta=\zeta_{0}+\zeta_{1}\frac{\dot{a}}{a}+\zeta_{2}\frac{\ddot{a}}{\dot{a}}$}
\end{table}

\subsubsection{Case 2: with $\zeta=\zeta_{0}+\zeta_{1}\frac{\dot{a}}{a}$}
\begin{figure}[tbp]
\centering
\includegraphics[width=.45\textwidth]{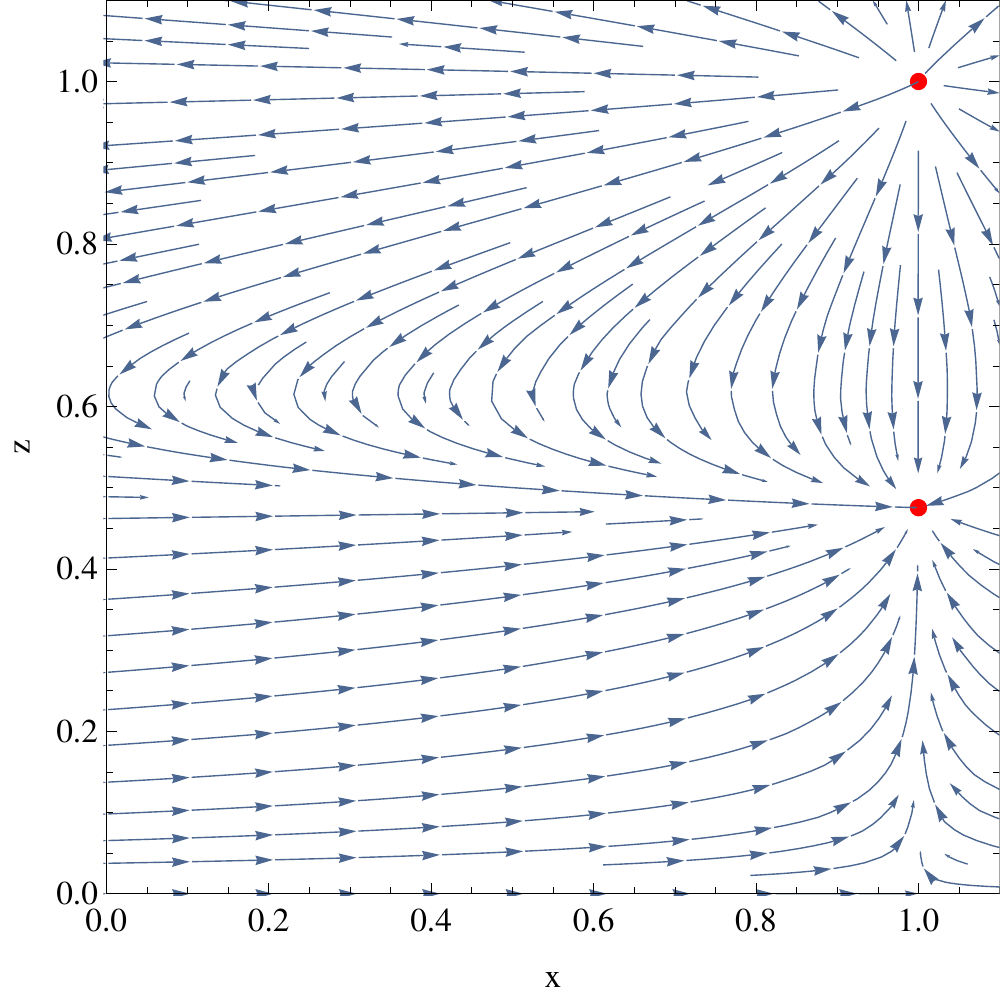}
\caption{\label{fig:2} The figure shows the phase space structure in
the $u-v$ plane corresponding to the Case 2
($\zeta=\zeta_{0}+\zeta_{1}\frac{\dot{a}}{a}$). The critical point
(1,1) in the upper right corner of the plot is a past attractor and
the point (1,0.475), below the first critical point, is a future
attractor. The direction of the trajectories is shown by the arrow
head.}
\end{figure}

Using the friedmann equations and conservation equation for matter
we get the autonomous equation for this case as,
\begin{equation}
\begin{aligned}
\begin{split}
u'&=\left(1-u\right)\left(\frac{\tilde{\zeta}_{0}}{v}+\tilde{\zeta}_{1}-\tilde{\zeta}_{0}\right)\\
v'&=\frac{1}{2}\left(1-v\right)\left(\tilde{\zeta}_{0}+\left(\tilde{\zeta}_{1}-\tilde{\zeta}_{0}-3\right)v\right).
\end{split}
\end{aligned}
\end{equation}
In this case also there are two critical points $(u_c,v_c)=(1,1)$
and $(1,\frac{\tilde{\zeta}_0}{\tilde{\zeta}_0-\tilde{\zeta}_1+3})$.
There properties are discussed below:
\begin{enumerate}
\item $(u_c,v_c)=(1,1)$ \\
This is a matter dominated solution representing the initial
singular state, since $v=1$ implies $H\rightarrow\infty$. The
critical point is same as that in case 1. The eigen values of the
corresponding Jacobian matrix are,
\begin{equation}
\lambda_1=\frac{3-\tilde{\zeta}_1}{2}=3.455,\ \ \ \
\lambda_2=-\tilde{\zeta}_1=3.91
\end{equation}
Since the eigen values are all positive, the critical point is an
unstable node or can be called as the past attractor. Thus it is a
source point of any orbit in the phase space. Using the values of
bulk viscous parameters from table ~\ref{tab:1}, we get $\omega=1.3$
and $q=2.45$ from Eqs. \eqref{equation1} and \eqref{deceleration1},
respectively. From these values it is clear that the point represent
a decelerated phase of the universe.
\item
$(u_c,v_c)=(1,\frac{\tilde{\zeta}_0}{\tilde{\zeta}_0-\tilde{\zeta}_1+3})=(1,0.475)$
\\ The eigen values of the corresponding Jacobian matrix are,
\begin{equation}
\lambda_1=\frac{\tilde{\zeta}_1-3}{2}=-3.455,\ \ \ \ \ \ \ \ \ \
\lambda_2=-3,
\end{equation}
using the value of $\tilde{\zeta}_1$ from Table ~\ref{tab:1}. Since
the eigen values are negative, this solution is a stable node and a
future attractor. So all trajectories in the phase space tends to
meet at this point. The equation of state parameter $\omega$ and the
deceleration parameter $q$ are both found to be -1, thereby
representing a de Sitter epoch.
\end{enumerate}
The phase plot of this case is shown in Fig. ~\ref{fig:2}. The phase
space trajectories starts from the critical point (1,1) and ends in
the point (1,0.475) in the u-v phase plane. The behaviour is same as
that in the first case. The results are summarized in table
~\ref{tab:3}.

\begin{table}[tbp]
\centering
\begin{tabular}{|c|l|c|p{1.7cm}|l|l|}
\hline
$(u_c,v_c)$ & $\lambda_1$ & $\lambda_2$ & Stability & $\omega$ & q  \\
\hline
$(1,1)$ & 3.45 & 3.9 & Unstable, Past attractor & 1.3 & 2.4 \\
$(1,0.475)$ & -3.45 & -3 & Stable, future attractor & -1 & -1 \\
\hline
\end{tabular}
\caption{\label{tab:3} Critical points for case 2, with
$\zeta=\zeta_{0}+\zeta_{1}\frac{\dot{a}}{a}$}
\end{table}

\subsubsection{Case 3: with $\zeta=\zeta_{0}$}
\begin{figure}[tbp]
\centering
\includegraphics[width=.45\textwidth]{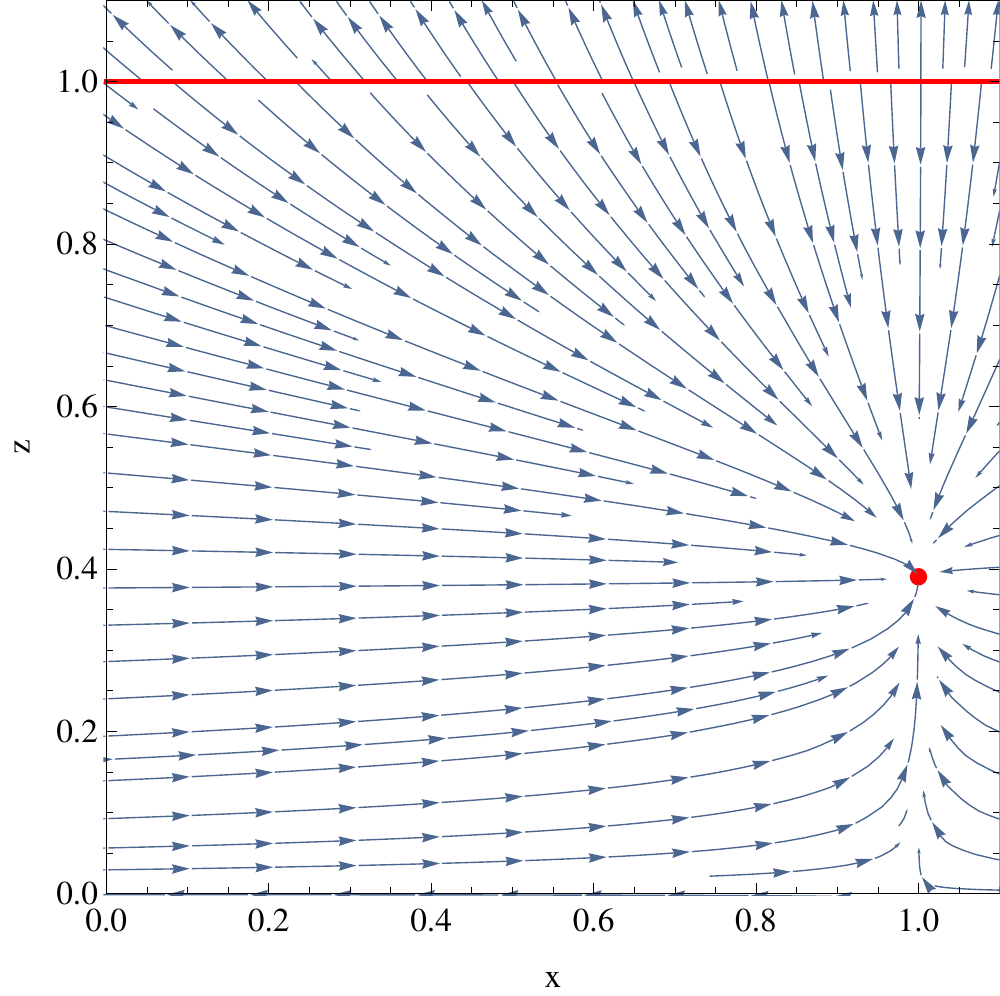}
\caption{\label{fig:3} The figure shows the phase space structure in
the $u-v$ plane corresponding to the Case 3 ($\zeta=\zeta_{0}$). The
direction of the trajectories is shown by the arrow head.}
\end{figure}

In this case, the autonomous equation reduces to,
\begin{equation}
\begin{aligned}
\begin{split}
u'&=\left(1-u\right)\left(\frac{\tilde{\zeta}_{0}}{v}-\tilde{\zeta}_{0}\right)\\
v'&=\frac{1}{2}\left(1-v\right)\left(\tilde{\zeta}_{0}-\left(\tilde{\zeta}_{0}+3\right)v\right).
\end{split}
\end{aligned}
\end{equation}
There exists two critical points:
\begin{enumerate}
\item $(u_c,v_c)=(u,1)$ \\
Here we see that the $u$ coordinate is variable, which can assume
any values ranging from 0 to 1, while $v$ coordinate is a constant
having value 1. As a result the critical point will not be an
isolated point (see Fig. ~\ref{fig:3}). It represents an initial
state of the universe since $H\rightarrow\infty$. The eigen values
of the corresponding Jacobian matrix are,
\begin{equation}
\lambda_1=\frac{3}{2}=1.5,\ \ \ \ \ \ \ \ \ \ \ \ \ \lambda_2=0
\end{equation}
Since these values are positive, it is unstable. The value of
equation of state parameter and the deceleration parameter are
$\omega=0$ and $q=0.5$. From these values it is clear that it
represents a matter dominated decelerated phase of the universe.
Unlike the other two cases, where the values of $\omega$ corresponds
to a stiff fluid, here in this case the value of $\omega$ indicates
the non-relativistic dark matter causing a usual decelerated phase.
\item
$(u_c,v_c)=(1,\frac{\tilde{\zeta}_0}{\tilde{\zeta}_0+3})=(1,0.39)$
\\
This corresponds to a matter dominated universe with
$\frac{H_0}{H}=1.564$, representing the future phase of the
universe. The eigen values are,
\begin{equation}\lambda_1=-\frac{3}{2}=-1.5,\ \ \ \ \ \ \ \ \ \ \ \
\ \ \lambda_2=-3
\end{equation}
The point is stable since both the eigen values are negative. The
values of the equation of state parameter and the deceleration
parameter are $\omega\sim-1$ and $q\sim-1$, which corresponds to a
de Sitter phase.
\end{enumerate}
The phase plot diagram is shown in Fig. ~\ref{fig:3}. From the
figure it is clear that the phase space trajectories originate from
the non-isolated critical point (or rather a critical line), which
is a past attractor (representing the matter dominated decelerated
epoch of the universe). These trajectories finally converge to the
stable critical point (1,0.39), representing the de-sitter phase.
This is similar to the behaviour of the $\Lambda$CDM model. The
results of the phase space analysis of the model are summarized in
table ~\ref{tab:4}.

\begin{table}[tbp]
\centering
\begin{tabular}{|c|l|c|p{4cm}|l|l|}
\hline
$(u_c,v_c)$ & $\lambda_1$ & $\lambda_2$ & Stability & $\omega$ & q  \\
\hline
$(u,1)$ & 1.5 & 0 & Unstable, Past attractor & 0 & 0.5 \\
$(1,0.39)$ & -1.5 & -3 & Stable, future attractor & -1 & -1 \\
\hline
\end{tabular}
\caption{\label{tab:4} Critical points for case 3, with
$\zeta=\zeta_{0}$}
\end{table}

\section{Universe with bulk viscous matter and radiation}
\label{sec:3} The realistic picture of the universe have a radiation
dominated phase followed by matter dominated epoch and a late
accelerated epoch. Inorder to know whether the bulk viscous model
predicts a prior radiation dominated phase, we study the phase space
structure of the model by including radiation as an additional
cosmic component. For such a universe, the friedmann equations
becomes,
\begin{equation}
 H^{2}=\frac{\rho_{m}+\rho_r}{3}
\end{equation}
\begin{equation}
\label{friedmann2} 2\dot{H}+3H^{2}=H
H_{0}\tilde{\zeta}-\frac{\rho_r}{3}
\end{equation}
The conservation equation for matter is given by Eq. \eqref{conser1}
and for radiation it is,
\begin{equation}
\label{conser2} \dot{\rho_{r}}+4H\rho_{r}=0.
\end{equation}
Eq. \eqref{friedmann2} can be modified using the radiation density
parameter $\Omega_r=\frac{\rho_{r}}{3H^{2}}$, then the derivative of
$H$ with respect to time $t$ becomes
\begin{equation}
\label{dH2} \dot{H}=\frac{1}{2}\left(H
H_{0}\tilde{\zeta}-3H^{2}-\Omega_{r}H^{2}\right)
\end{equation}
Substituting this in Eq. \eqref{zeta}, the total dimensionless bulk
viscous parameter $\tilde{\zeta}$ takes the form,
\begin{equation}
\label{zeta2}
\tilde{\zeta}=\frac{1}{2-\tilde{\zeta}_{2}}\left[2\tilde{\zeta}_{0}+\left(2\tilde{\zeta}_{1}-\tilde{\zeta}_{2}-\tilde{\zeta}_{2}\Omega_{r}\right)
\frac{H}{H_{0}}\right],
\end{equation}
which will reduces to Eq. \eqref{zeta1} for $\Omega_r=0$. The
deceleration parameter $q$ and equation of state parameter $\omega$
can be obtained by substituting Eqs. \eqref{dH2} and \eqref{zeta2}
in Eqs. \eqref{deceleration} and \eqref{equation} as
\begin{equation}
\label{eq:decel}
q=\frac{1}{2-\tilde{\zeta}_{2}}\left(1-\tilde{\zeta}_1+\Omega_r-\tilde{\zeta}_0(\frac{H_0}{H})\right)
\end{equation}
\begin{equation}
\label{eq:eqn}
\omega=\frac{1}{3(2-\tilde{\zeta}_{2})}\left(2\Omega_r-2\tilde{\zeta}_1+\tilde{\zeta}_{2}-2\tilde{\zeta}_0(\frac{H_0}{H})\right]
\end{equation}
which are reducing to Eqs. \eqref{decelerationa} and
\eqref{equationa} for $\Omega_r=0$. In the radiation dominated case
(i.e., when $\Omega_r\rightarrow1$, then
$\frac{H_0}{H}\rightarrow0$), the deceleration parameter and the
equation of state reduces to,
\begin{equation}
q\sim\frac{2-\tilde{\zeta}_1}{2-\tilde{\zeta}_{2}}
\end{equation}
\begin{equation}
\omega\sim\frac{2-2\tilde{\zeta}_1+\tilde{\zeta}_2}{3(2-\tilde{\zeta}_{2})}
\end{equation}
When radiation is the dominant component,universe there would be no
acceleration in expansion such that $q>0$ and $\omega>-\frac{1}{3}$.
These conditions constrains the bulk viscous parameters as
$\tilde{\zeta}_1<2$ and $\tilde{\zeta}_2<2$. In the extreme limit
corresponding to the radiation dominated phase, $q=1$ and
$\omega=\frac{1}{3}$ and is corresponding to
$\tilde{\zeta}_1=\tilde{\zeta}_2$.
\subsection{Phase space analysis}
For doing the phase space analysis, we are defining the phase space
co-ordinates as
\begin{equation}
\begin{aligned}
\begin{split}
u&=\Omega_m=\frac{\rho_m}{3H^2} \\
y&=\Omega_r=\frac{\rho_r}{3H^2}\\
v&=\frac{1}{\frac{H_0}{H}+1}
\end{split}
\end{aligned}
\end{equation}
Contrary to the previous discussion, here we take $\zeta=\zeta_{0}$
as Case 1,$\zeta=\zeta_{0}+\zeta_{1}\frac{\dot{a}}{a}$ as Case 2 and
$\zeta=\zeta_{0}+\zeta_{1}\frac{\dot{a}}{a}+\zeta_{2}\frac{\ddot{a}}{\dot{a}}$
as Case 3.
\subsubsection{Case 1: with
$\zeta=\zeta_{0}$} Using Eqs. \eqref{conser1}, \eqref{conser2} and
\eqref{dH2}, the autonomous equations satisfied by the phase space
co-ordinates becomes,
\begin{equation}
\begin{aligned}
\begin{split}
u'&=\tilde{\zeta}_0(1-u)(\frac{1-v}{v})+uy \\
y'&=\frac{y}{v}\left(\tilde{\zeta}_0(v-1)+v(y-1)\right) \\
v'&=\frac{(1-v)}{2}\left(\tilde{\zeta}_0(1-v)-v(y+3)\right)
\end{split}
\end{aligned}
\end{equation}
The Jacobian matrix for this can be obtained from Eq.
\eqref{eq:matrix1} by setting $\tilde{\zeta}_1=\tilde{\zeta}_2=0$.
The critical points are,
\begin{enumerate}
  \item
  $(u_c,y_c,v_c)=(0,1,1)$ \\
  This corresponds to the radiation dominated phase of the universe. The eigen values of the corresponding Jacobian matrix are,
  \begin{equation}
  \lambda_1=1, \ \ \
  \lambda_2=1,\ \ \ \lambda_3=2.
  \end{equation}
  All the eigen values are positive, indicating an unstable node (past attractor). The equation of
  state parameter and the deceleration parameter corresponding to this critical point
  can be obtained by substituting the values of $u_c$, $y_c$ and $v_c$ in Eqs. \eqref{eq:eqn} and \eqref{eq:decel},
  respectively, and are found to be $\omega=\frac{1}{3}$ and
  $q=1$. These values confirms that the point is the radiation
  dominated phase of the universe.
  \item $(u_c,y_c,v_c)=(u,0,1)$ \\
  The eigen values of the corresponding Jacobian matrix are,
  \begin{equation}
  \lambda_1=-1, \ \ \
  \lambda_2=0,\ \ \ \lambda_3=\frac{3}{2}
  \end{equation}
  These values shows that the point is a saddle point. The equation of
  state parameter and deceleration parameter are found to be,
  $\omega=0$ and
  $q=\frac{1}{2}$ respectively from Eqs. \eqref{eq:eqn} and \eqref{eq:decel}, indicating that the universe is matter dominated without accelerating, hence
  $u_c\sim1$.
  \item
  $(u_c,y_c,v_c)=(1,0,\frac{\tilde{\zeta}_0}{\tilde{\zeta}_0+3})$ \\
  The eigen values of the corresponding Jacobian matrix are,
  \begin{equation}
  \lambda_1=-4,\ \ \ \lambda_2=-3, \ \ \
  \lambda_3=-\frac{3}{2}
  \end{equation}
  The critical point is a stable node, since all the eigen values are negative.From Eqs. \eqref{eq:eqn} and \eqref{eq:decel}, the equation of
  state parameter, $\omega=-1$ and deceleration parameter
  $q=-1$, independent of the value of $\tilde{\zeta}_0$. This represent a de Sitter type phase.
\end{enumerate}
Thus this case predicts a universe beginning with a radiation
dominated phase (past attractor) and then transit to a decelerated
matter dominated phase (saddle point) and then finally evolving to a
de Sitter type universe (stable future attractor). Thus it has a
close resemblance with the conventional evolution of the universe.
The results are summarized in Table ~\ref{tab:7}.
\begin{table}[tbp]
\centering
\begin{tabular}{|l|l|c|c|c|c|c|}
\hline
$(u_c,y_c,v_c)$ & $\lambda_1$ & $\lambda_2$ & $\lambda_3$ & Stability & $\omega$ & q  \\
\hline
$(0,1,1)$ & 1 & 1 & 2 & Unstable node & $\frac{1}{3}$ & 1 \\
$(u,0,1)$ & -1 & 0 & $\frac{3}{2}$ & Saddle & 0 & $\frac{1}{2}$ \\
$(1,0,\frac{\tilde{\zeta}_{0}}{\tilde{\zeta}_{0}+3)})$ & -4 & -3 & $-\frac{3}{2}$ & Stable node & -1 & -1 \\
\hline
\end{tabular}
\caption{\label{tab:7} Critical points for Case 1: with
$\zeta=\zeta_{0}$}
\end{table}

\subsubsection{Case 2: with
$\zeta=\zeta_{0}+\zeta_{1}\frac{\dot{a}}{a}$} In this case the
autonomous equations are,
\begin{equation}
\begin{aligned}
\begin{split}
u'&=\frac{\tilde{\zeta}_0(1-u)}{v}+(1-u)(\tilde{\zeta}_1-\tilde{\zeta}_0)+uy \\
y'&=y\left(\frac{\tilde{\zeta}_0(v-1)}{v}+y-1-\tilde{\zeta}_1\right)\\
v'&=\frac{1}{2-\tilde{\zeta}_{2}}\left(1-v\right)\left(\tilde{\zeta}_0(1-v)+(\tilde{\zeta}_1-3-y)v\right)
\end{split}
\end{aligned}
\end{equation}
The Jacobian matrix for this autonomous system is given by Eq.
\eqref{eq:matrix1} provided $\tilde{\zeta}_2=0$. The critical points
are,
\begin{enumerate}
  \item
  $(u_c,y_c,v_c)=(-\tilde{\zeta}_1,\tilde{\zeta}_1+1,1)$ \\
  The eigen values of the corresponding Jacobian matrix are,
  \begin{equation}
  \lambda_1=\tilde{\zeta}_1+1, \
  \lambda_2=1, \ \lambda_3=2.
  \end{equation}
  The point will be unstable node if $\tilde{\zeta}_1>-1$, otherwise it will be a saddle point. The equation of
  state parameter, $\omega=\frac{1}{3}$ and deceleration parameter
  $q=1$, independent of $\tilde{\zeta}_0$ and $\tilde{\zeta}_1$. Hence if $\tilde{\zeta}_1=0$,
  the point will indicate an exact radiation dominated universe.
  \item $(u_c,y_c,v_c)=(1,0,1)$ \\
  This corresponds to a matter dominated initial stage of the universe. The eigen values of
  the corresponding Jacobian matrix are,
  \begin{equation}
  \lambda_1=-(\tilde{\zeta}_1+1), \ \lambda_2=-\tilde{\zeta}_1, \
  \lambda_3=\frac{3-\tilde{\zeta}_1}{2}.
  \end{equation}
  The point will be unstable node if $\tilde{\zeta}_1<-1$, a stable one if $\tilde{\zeta}_1>3$ and a saddle point otherwise.
  For this point, the equation of
  state, $\omega=-\frac{\tilde{\zeta}_1}{3}$ and the deceleration parameter
  $q=\frac{1-\tilde{\zeta}_1}{2}$. If $\tilde{\zeta}_1<-1$, then the values of $\omega$ and $q$ will
  not represent a conventional matter dominated universe. So only if $\tilde{\zeta}_1=0$, this will
  represent a conventional matter dominated universe without acceleration.
  \item
  $(u_c,y_c,v_c)=(1,0,\frac{\tilde{\zeta}_0}{\tilde{\zeta}_0-(\tilde{\zeta}_1-3)})$
  \\
  The eigen values are,
  \begin{equation}
  \lambda_1=-4, \ \lambda_2=-3, \
  \lambda_3=\frac{1}{2}(\tilde{\zeta}_1-3).
  \end{equation}
  The point will be a stable one if $\tilde{\zeta}_1<3$, otherwise it will be a saddle point. The equation of
  state parameter, $\omega=-1$ and deceleration parameter
  $q=-1$, independent of the values of viscous parameters, representing a de Sitter type universe. So If
  $\tilde{\zeta}_1=0$, the point will be represent a stable future
  attractor with same values of $\omega$ and $q$.
\end{enumerate}
In order to represent a realistic picture, the first critical point
must be a unstable (past attractor) radiation dominated phase, the
second must be a matter dominated phase without acceleration (saddle
point) and the third must corresponds to the stable accelerated
phase of the universe. Inorder to satisfy this, $\tilde{\zeta}_1$
should be equal to zero. The results of the analysis is given in
Table ~\ref{tab:6}
\begin{table}[tbp]
\centering
\begin{tabular}{|l|c|c|c|p{4cm}|c|c|}
\hline
$(u_c,y_c,v_c)$ & $\lambda_1$ & $\lambda_2$ & $\lambda_3$ & Stability & $\omega$ & q  \\
\hline
$(-\tilde{\zeta}_1, \tilde{\zeta}_1+1, 1)$ & $\tilde{\zeta}_1+1$ & 1 & 2 & Unstable node if $\tilde{\zeta}_1>-1$, otherwise a saddle point & $\frac{1}{3}$ & 1 \\
$(1,0,1)$ & $-(\tilde{\zeta}_1+1)$ & $-\tilde{\zeta}_1$ & $\frac{3-\tilde{\zeta}_1}{2}$ & Unstable node if $\tilde{\zeta}_1<-1$, Stable if $\tilde{\zeta}_1>3$, a saddle point otherwise & $\frac{-\tilde{\zeta}_1}{3}$ & $\frac{1-\tilde{\zeta}_1}{2}$ \\
$(1, 0, \frac{\tilde{\zeta}_{0}}{\tilde{\zeta}_{0}-(\tilde{\zeta}_1-3)})$ & -4 & -3 & $\frac{1}{2}(\tilde{\zeta}_1-3)$ & Unstable node if $\tilde{\zeta}_1<3$, otherwise a saddle point & -1 & -1 \\
\hline
\end{tabular}
\caption{\label{tab:6} Critical points for Case 2: with
$\zeta=\zeta_{0}+\zeta_{1}\frac{\dot{a}}{a}$}
\end{table}

\subsubsection{Case 3: with
$\zeta=\zeta_{0}+\zeta_{1}\frac{\dot{a}}{a}+\zeta_{2}\frac{\ddot{a}}{\dot{a}}$}
In this case the phase space variables satisfy the autonomous
equations,
\begin{equation}
\begin{aligned}
\begin{split}
u'=&\frac{1}{v(2-\tilde{\zeta}_{2})}(2\tilde{\zeta}_0(1-u)(1-v)+\\ & v((u-y-1)\tilde{\zeta}_{2}+2(1-u)\tilde{\zeta}_1+2yu))\\
y'=&\frac{1}{v(2-\tilde{\zeta}_{2})}2y\left(\tilde{\zeta}_0(v-1)+(y-1-\tilde{\zeta}_1+\tilde{\zeta}_{2})v\right)\\
v'=&\frac{1}{2-\tilde{\zeta}_{2}}\left(1-v\right)\left(\tilde{\zeta}_0(1-v)+(\tilde{\zeta}_1+\tilde{\zeta}_{2}-3-y)v\right)
\end{split}
\end{aligned}
\end{equation}

\begin{table}[tbp]
\centering
\begin{tabular}{|p{4cm}|c|c|c|p{3.5cm}|c|c|}
\hline
$(u_c,y_c,v_c)$ & $\lambda_1$ & $\lambda_2$ & $\lambda_3$ & Stability & $\omega$ & q  \\
\hline
$(\tilde{\zeta}_2-\tilde{\zeta}_1,1-(\tilde{\zeta}_2-\tilde{\zeta}_1,1)$ & $\frac{-2(\tilde{\zeta}_1-\tilde{\zeta}_2+1)}{\tilde{\zeta}_2-2}$ & 1 & 2 & Unstable node if $\tilde{\zeta}_2<2$, Saddle point otherwise & $\frac{1}{3}$ & 1 \\
$(1,0,1)$ &
$\frac{2(\tilde{\zeta}_1-\tilde{\zeta}_2+1)}{\tilde{\zeta}_2-2}$ &
$\frac{2\tilde{\zeta}_1-\tilde{\zeta}_2}{\tilde{\zeta}_2-2}$ &
$\frac{\tilde{\zeta}_1+\tilde{\zeta}_2-3}{\tilde{\zeta}_2-2}$ &
Unstable node if (i)
$\tilde{\zeta}_2>2,\tilde{\zeta}_1>1,\tilde{\zeta}_1-\tilde{\zeta}_2>-1$,
(ii)
$\tilde{\zeta}_2<2,\tilde{\zeta}_1<1,\tilde{\zeta}_1-\tilde{\zeta}_2<-1$
& $\frac{-2\tilde{\zeta}_1+\tilde{\zeta}_2}{3(2-\tilde{\zeta}_2)}$ &
$\frac{1-\tilde{\zeta}_1}{2-\tilde{\zeta}_2}$
\footnotemark[1]\\
$(1,0,\frac{\tilde{\zeta}_{0}}{\tilde{\zeta}_{0}-(\tilde{\zeta}_1+\tilde{\zeta}_2-3)})$ & -4 & -3 & $-\frac{\tilde{\zeta}_1+\tilde{\zeta}_2-3}{\tilde{\zeta}_2-2}$ & Stable node if (i) $\tilde{\zeta}_0>0, \tilde{\zeta}_2<2$, (ii) $\tilde{\zeta}_0<0,\tilde{\zeta}_2>2$, Saddle point otherwise & -1 & -1 \\
\hline
\end{tabular}
\caption{\label{tab:5} Critical points for Case 3: with
$\zeta=\zeta_{0}+\zeta_{1}\frac{\dot{a}}{a}+\zeta_{2}\frac{\ddot{a}}{\dot{a}}$}
\footnotemark[1]{When $\tilde{\zeta}_1=\tilde{\zeta}_2=0$, the
critical point corresponds to $w=0$, $q=\frac{1}{2}$ implying a
matter dominated universe without acceleration.}
\end{table}

The Jacobian matrix for these set of autonomous equation is given
by,

\begin{equation}
\label{eq:matrix1}
\begin{bmatrix}
\left(\frac{2\tilde{\zeta}_0(1-v)+(2\tilde{\zeta}_1-\tilde{\zeta}_2-2y)v}{(\tilde{\zeta}_2-2)v}\right)_0
& \left(\frac{\tilde{\zeta}_2-2u}{\tilde{\zeta}_2-2}\right)_0 & \left(\frac{2\tilde{\zeta}_0(1-u)}{(\tilde{\zeta}_2-2)v^2}\right)_0\\
0 &
\left(\frac{2(\tilde{\zeta}_0-\tilde{\zeta}_0v+(1+\tilde{\zeta}_1-\tilde{\zeta}_2-2y)v)}{(\tilde{\zeta}_2-2)v}\right)_0
& \left(\frac{-2\tilde{\zeta}_0y}{(\tilde{\zeta}_2-2)v^2}\right)_0\\
0 & \left(\frac{(1-v)v}{\tilde{\zeta}_2-2}\right)_0 &
\left(\frac{2\tilde{\zeta}_0(1-v)+(\tilde{\zeta}_1+\tilde{\zeta}_2-3-y)(2v-1)}{\tilde{\zeta}_2-2}\right)_0
\end{bmatrix}
\end{equation}

The critical points $(u_c,y_c,v_c)$ of these equations are
\begin{enumerate}
  \item
$(u_c,y_c,v_c)=(\tilde{\zeta}_2-\tilde{\zeta}_1,1-(\tilde{\zeta}_2-\tilde{\zeta}_1),1)$\\
  For this solution to represent the realistic phase (for example, matter dominated or radiation dominated) of the
  universe, the bulk viscous parameters should satisfy
  the condition, $0\leq\tilde{\zeta}_2-\tilde{\zeta}_1\leq1$.
  The eigen values of the corresponding Jacobin matrix are,
  \begin{equation}
  \label{eigen1}
  \lambda_1=\frac{-2(\tilde{\zeta}_1-\tilde{\zeta}_2+1)}{\tilde{\zeta}_2-2},\
  \ \
  \lambda_2=1, \ \ \ \lambda_3=2.
  \end{equation}
  The point will be unstable if $\tilde{\zeta}_2-2<0$ and a saddle
  point otherwise.
  The equation of
  state parameter and deceleration parameter corresponding to this critical point are,
  $\omega=\frac{1}{3}$ and $q=1$.
  \item $(u_c,y_c,v_c)=(1,0,1)$\\
  This point corresponds to a matter dominated phase of the universe. The eigen values of the corresponding jacobian matrix are,
  \begin{equation}
  \label{eigen2}
  \lambda_1=\frac{2(\tilde{\zeta}_1-\tilde{\zeta}_2+1)}{\tilde{\zeta}_2-2},
  \lambda_2=\frac{2\tilde{\zeta}_1-\tilde{\zeta}_2}{\tilde{\zeta}_2-2},
  \lambda_3=\frac{\tilde{\zeta}_1+\tilde{\zeta}_2-3}{\tilde{\zeta}_2-2}.
  \end{equation}
  This critical point will be unstable if (i) $\tilde{\zeta}_2>2, \tilde{\zeta}_1>1, \tilde{\zeta}_1-\tilde{\zeta}_2>-1$
  or if
  (ii) $\tilde{\zeta}_2<2, \tilde{\zeta}_1<1, \tilde{\zeta}_1-\tilde{\zeta}_2<-1$.
  The equation of state parameter and the deceleration parameter corresponding to this critical point are,
  $\omega=\frac{2\tilde{\zeta}_1-\tilde{\zeta}_2}{3(\tilde{\zeta}_2-2)}$ and
  $q=\frac{\tilde{\zeta}_1-1}{\tilde{\zeta}_2-2}$.
  \item
 $(u_c,y_c,v_c)=(1,0,\frac{\tilde{\zeta}_0}{\tilde{\zeta}_0-(\tilde{\zeta}_1+\tilde{\zeta}_2-3)})$\\
  This also represents a matter dominated universe with $\frac{H_0}{H}=\frac{3-(\tilde{\zeta}_1+\tilde{\zeta}_2)}{\tilde{\zeta}_0}$.
  The eigen values of the Jacobian matrix are,
  \begin{equation}
  \lambda_1=-4,\ \ \ \lambda_2=-3, \ \ \
  \lambda_3=-\frac{\tilde{\zeta}_1+\tilde{\zeta}_2-3}{\tilde{\zeta}_2-2}.
  \end{equation}
  The relation $\frac{H_0}{H}=\frac{3-(\tilde{\zeta}_1+\tilde{\zeta}_2)}{\tilde{\zeta}_0}>0$ holds
  if $\tilde{\zeta}_0>0$ and $\tilde{\zeta}_1+\tilde{\zeta}_2<3$
  or if $\tilde{\zeta}_0<0$ and $\tilde{\zeta}_1+\tilde{\zeta}_2>3$. Applying this condition to the eigen value $\lambda_3$
  we find that this critical point will be stable (or future attractor) if (i) $\tilde{\zeta}_2<2$ for $\tilde{\zeta}_0>0$ or if (ii)
  $\tilde{\zeta}_2>2$ for  $\tilde{\zeta}_0<0$. It is a saddle point otherwise.
  The equation of
  state parameter, $\omega=-1$ and deceleration parameter
  $q=-1$ implies a
  de sitter like universe.
\end{enumerate}
The above critical points would represent the realistic evolution of
the universe, if, successively, the first critical point is a
radiation dominated one, the second one is a matter dominated phase
without acceleration and the last one be a matter dominated phase
with acceleration. For this, first of all the values of the viscous
coefficients must be such that $\tilde{\zeta}_1\sim\tilde{\zeta}_2$.
Under this conditions the critical points becomes
\begin{enumerate}
  \item $(u_c,y_c,v_c)=(0,1,1)$, corresponding to radiation
dominated phase with $\omega=\frac{1}{3}$ and $q=1$
  \item $(u_c,y_c,v_c)=(1,0,1)$, corresponding to matter
dominated phase with
$\omega=\frac{\tilde{\zeta}_2}{3(\tilde{\zeta}_2-2)}$ and
$q=\frac{\tilde{\zeta}_2-1}{\tilde{\zeta}_2-2}$
  \item
$(u_c,y_c,v_c)=(1,0,\frac{\tilde{\zeta}_0}{\tilde{\zeta}_0-(2\tilde{\zeta}_2-3)})$,
corresponding to accelerating phase with $\omega=-1$ and $q=-1$.
\end{enumerate}
Then by analyzing the eigen values, it is found that the first
critical point (Eq.\eqref{eigen1}) will be a past attractor if
$\tilde{\zeta}_2<2$. Under this condition, the second critical
point, given by Eq. \eqref{eigen2}, corresponding to the matter
dominated phase, will be a saddle point. The third critical point,
corresponding to the accelerated phase, will be a stable one if
$\tilde{\zeta}_2<\frac{3}{2}$. However, under this condition, there
is a chance for $\omega$ and $q$ to become negative in the case of
matter dominated phase corresponding to second set of critical
points \eqref{eigen2}. This doesn't represent a conventional matter
dominated phase of the universe. For this critical point to
represent a matter dominated phase without acceleration, it requires
$\omega=0$ and $q=\frac{1}{2}$. This is possible only if
$\tilde{\zeta}_1\simeq\tilde{\zeta}_2=0$. Due to this conditions,
the nature of the first and the last critical points will not be
affected. i.e., the first critical point will be a radiation
dominated past attractor, the second one will be an unaccelerated
matter dominated saddle point and the third will be a stable node
corresponding to a de Sitter phase. Thus it predicts a universe
starting from a radiation dominated era and then entering a
decelerated matter dominated phase and then finally evolving to the
de Sitter universe.Thus we see that unless
$\tilde{\zeta}_1=\tilde{\zeta}_2=0$, the model doesn't predict a
prior radiation dominated phase  and a decelerated matter dominated
phase of the universe. The results are summarized in Table
~\ref{tab:5}

\section{Conclusion}
\label{sec:4} We have done the phase space analysis of the universe
with bulk viscous matter having bulk viscous coefficient of the form
(i)$\zeta=\zeta_{0}$, (ii)
$\zeta=\zeta_{0}+\zeta_{1}\frac{\dot{a}}{a}$, (iii)
$\zeta=\zeta_{0}+\zeta_{1}\frac{\dot{a}}{a}+\zeta_{2}\frac{\ddot{a}}{\dot{a}}$.
First we have considered a universe containing bulk viscous matter
alone and for the above cases, the value of the bulk viscous
parameters are extracted using SCP "Union" Type I a Supernova data.
Using these the asymptotic properties of the model are studied and
phase space for each cases are plotted. It is found that for all the
three cases, it predicts a prior matter dominated universe without
acceleration, which is an unstable node and a stable matter
dominated universe with acceleration, similar to the de Sitter
phase.

Secondly we consider a universe including radiation and bulk viscous
matter to check whether it predicts a prior conventional radiation
dominance also. The phase space analysis of the model is done for
the three different cases separately. For the case
$\zeta=\zeta_{0}$, it is found that it predicts a universe beginning
with a radiation dominated phase (past attractor) and then transit
to a decelerated matter dominated phase (saddle point) and then
finally evolving to a de Sitter type universe (stable future
attractor). The other two cases, i.e., with
$\zeta=\zeta_{0}+\zeta_{1}\frac{\dot{a}}{a}$ and
$\zeta=\zeta_{0}+\zeta_{1}\frac{\dot{a}}{a}+\zeta_{2}\frac{\ddot{a}}{\dot{a}}$,
predicts a stable accelerated phase of the universe similar to a de
Sitter type with $\omega=-1$ and $q=-1$. However, these doesn't
predicts a prior radiation dominated phase and conventional
decelerated matter dominated phase of the universe, unless
$\zeta_{1}=\zeta_{2}=0$. There are approaches in the literature
~\cite{fabris1,Chimento1997,Colistete2007} where bulk viscosity is
included through $\zeta=\alpha\rho^m$, but there is no guarantee
that this approach will give the same result compared to the present
case, for instant in reference ~\cite{Velton2011} it is shown that
these two approaches are different in the structure formation

Thus, the bulk viscous model, which is an alternative to dark
energy, will predicts all the conventional phases and evolution of
the universe only with constant bulk viscous coefficient,
$\zeta=\zeta_0$. For the other two cases, with
$\zeta=\zeta_{0}+\zeta_{1}\frac{\dot{a}}{a}$ and
$\zeta=\zeta_{0}+\zeta_{1}\frac{\dot{a}}{a}+\zeta_{2}\frac{\ddot{a}}{\dot{a}}$,
even though explains the late acceleration of the universe
~\cite{athira}, but fails to predict the prior radiation dominated
phase and the decelerated matter dominated phase of the universe.\\[0.1in]

{\bf Acknowledgment}\\
 The author
(AS) is thankful to DST for giving financial support through INSPIRE
fellowship. Author (TKM) is thankful to The Inter-University Centre
for Astronomy and Astrophysics (IUCAA), Pune, India for the
hospitality during the visits where part of the work has been
carried out. \\[0.1in]

{\bf Appendix}\\
\begin{enumerate}
\item Validity of Generalised second law for the present model\\
In the FLRW space-time, the law of generation of the local entropy
is given as \cite{weinberg2}
\begin{equation}
\label{entropy}
T\nabla_{\nu}s^{\nu}=\zeta(\nabla_{\nu}u^{\nu})^{2}=9H^{2}\zeta
\end{equation}
where $T$ is the temperature and $\nabla_{\nu}s^{\nu}$ is the rate
of generation of entropy in unit volume. The second law of
thermodynamics will be satisfied if,
\begin{equation}
T\nabla_{\nu}s^{\nu}\geq 0
\end{equation}
which implies from equation (\ref{entropy}) that
\begin{equation}
\label{en} \zeta\geq 0.
\end{equation}

For
$\zeta=\zeta_{0}+\zeta_{1}\frac{\dot{a}}{a}+\zeta_{2}\frac{\ddot{a}}{\dot{a}}$,
it is found that the total $\zeta$ is negative when $z>0.8$
\cite{athira}, thereby violating local second law in the early
universe. But when $\zeta=\zeta_0$, $\zeta$ always remains positive
through out the evolution of the universe (since $\zeta_0>0$) and
hence satisfying the local second law of thermodynamics.

However, if one consider the Generalised second law (GSL) which
includes the total entropy of the universe plus that of the horizon,
it is found that total entropy is always on the increase for total
$\zeta$, if apparent horizon is considered as the boundary
\cite{athira}. On taking account of the validity of GSL, it can be
reasonably argued that in the early universe where $\zeta$ becomes
negative, the total pressure becomes positive and the viscous matter
will act as an ordinary non-relativistic matter causing decelerated
expansion. There are conventional dark energy models which act as
the non-relativistic matter in the early phases causing decelerated
expansion \cite{Bento,Andrea,Gorini,Zimdahl}. There are also works
\cite{Brevik,Radi1,Radi2,Li}, showing that the entropy change can
become negative depending on the equation of state of matter. In the
current literature, there are publications dealing the problems with
negative viscous coefficients \cite{Brevik}. In this reference the
authors has point out that the positivity of $\zeta$ in conventional
cosmology is based upon the requirement that the change of entropy
in a non-equilibrium system is positive, and they argues that the
possibility of allowing for negative values of $\zeta$ is not so
unreasonable in view of the general bizarre properties of the dark
energy fluid, as far as temperature is positive. Apart from this our
model also satisfies GSL with total $\zeta$.

Now we will consider event horizon as the boundary for analyzing the
validity of GSL for $\zeta=\zeta_0$.
 The radius of the event horizon is given as,
\begin{equation}
R_E=a\int_a^\infty\frac{da}{Ha^2}
\end{equation}
where $a$ is the scale factor and $H$ is the Hubble parameter given
as \cite{athira},
\begin{equation}
\label{hubbleina}
H(a)=H_{0}\left[a^{-\frac{3}{2}}\left(1-\frac{\tilde{\zeta}_{0}}{3}\right)+\frac{\tilde{\zeta}_{0}}{3}\right]
\end{equation}
$H_0$ is the present value of the Hubble parameter. The radius of
the event horizon then obtained as,
\begin{equation}
\begin{aligned}
\begin{split}
R_E=\frac{a}{H_0}(4.16+ 2.65\arctan[0.58-
1.4\sqrt{a}]-1.53\log[0.97+1.18\sqrt{a}]\\+0.76\log[0.94-1.14\sqrt{a}+1.38
a])
\end{split}
\end{aligned}
\end{equation}
The entropy associated with the event horizon is
\begin{equation}
S_E=\frac{A}{4}
\end{equation}
$A=4\pi {R_E}^2$ is the area of the horizon. So entropy becomes
\begin{equation}
S_E=\pi {R_E}^2
\end{equation}
The temperature of the event horizon can be defined as
$T_E=\frac{1}{2\pi R_E}$. Using these we get\cite{karami},
\begin{equation}
\label{horizonentropy} T_E\dot{S}_E=\dot{R}_E=HR_E-1
\end{equation}
The entropy of matter can be obtained using the Gibb's relation,
\begin{equation}
\begin{aligned}
\begin{split}
\label{gibbs} T_{m}dS_{m}=&d(\rho_{m}V)+PdV
\\=&(\rho_m+P)dV+V d\rho_m
\end{split}
\end{aligned}
\end{equation}
where $T_{m}$ is the temperature of the bulk viscous matter,
$V=\frac{4}{3}\pi R_{E}^{3}$ is the volume enclosed by the event
horizon. Using the expression for pressure $P=-3H\zeta_0=H
H_0\tilde{\zeta}_0$ ($\tilde{\zeta}_0=\frac{3\zeta}{H_0}$ is the
dimensionless viscous parameter), the conservation equation and the
relation $\dot{R}_E=HR_E-1$ , we get
\begin{equation}
\label{matterentropy} T_m\dot{S}_m=4\pi
R_E^2(HH_0\tilde{\zeta_0}-3H^2)
\end{equation}

Under equilibrium conditions, the temperature $T_{m}$ of the viscous
matter and that of the horizon $T_{E}$ are equal, $T_{m}=T_{E}=T$.
Adding Eqs. (\ref{horizonentropy}) and (\ref{matterentropy}), we
get,
\begin{equation}
\label{gsl} T(\dot{S}_E+\dot{S}_m)=4\pi
R_E^2(HH_0\tilde{\zeta_0}-3H^2)+HR_E-1
\end{equation}
For GSL to be valid, $T(\dot{S}_E+\dot{S}_m)>0$. We have check the
validity by numerically plotting Eq. (\ref{gsl}) with respect to $a$
and is shown in the figure \ref{fig:4}.

\begin{figure}[tbp]
\centering
\includegraphics[width=.45\textwidth]{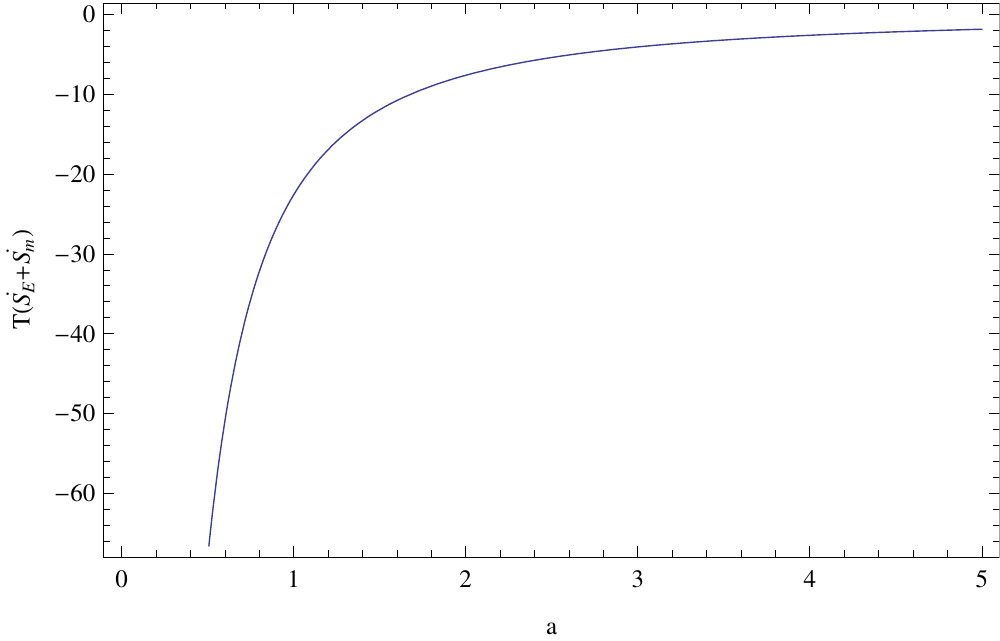}
\caption{\label{fig:4} Plot of $T(\dot{S}_E+\dot{S}_m)$ with respect
to the scale factor $a$.}
\end{figure}

The plot shows that the GSL is violated when we take event horizon
as the boundary. So, in our model GSL is satisfied at the apparent
horizon but violated at the event horizon. At this juncture, one may
note that a more novel GSL was proposed by Bousso et.al
\cite{Bousso} regardless of whether an event horizon is present.
However, the validity of this new GSL is to checked for our model

There are many works in literature in tune with our result regarding
the validity of GSL. In references \cite{Zhov,Wang}, the authors
have shown that in general, for an accelerating universe, GSL of
thermodynamics holds only in the case where the enveloping surface
is the apparent horizon, but not in the case of the event horizon.
There are also many other dark energy models which shows the same
behavior. Some models are viscous model \cite{Setare,Akbar},
interacting dark energy model with dark matter \cite{karami},
Holographic Ricci dark energy model \cite{Titus,Praseetha}, DGP
model \cite{Sheykhi1}, braneworld model \cite{Sheykhi2}. In all
these references, it is found that the event horizon in an
accelerating universe is not a boundary from the thermodynamical
point of view. In lieu of these, apparent horizon can be considered
as the proper thermodynamic boundary.

\item Statefinder parameter diagnostic for $\zeta=\zeta_0$ (comparison with $\Lambda$CDM model)\\
 For
comparison we have make use of the statefinder parameter diagnostic
introduced by Sahni et al \cite{Sahni}. The statefinder parameters
$\{r,s\}$ are defined as,
\begin{equation}
r=\frac{\dddot{a}}{aH^3}, \ \ \ \ \ \
s=\frac{r-1}{3\left(q-\frac{1}{2}\right)}.
\end{equation}
In terms of $h=\frac{H}{H_0}$, $r$ and $s$ can be written as
\begin{equation}
r=\frac{1}{2h^{2}}\frac{d^{2}h^2}{dx^2}+\frac{3}{2h^2}\frac{dh^2}{dx}+1
\end{equation}
\begin{equation}
s=-\frac{\frac{1}{2h^{2}}\frac{d^{2}h^2}{dx^2}+\frac{3}{2h^2}\frac{dh^2}{dx}}{\frac{3}{2h^2}\frac{dh^2}{dx}+\frac{9}{2}}.
\end{equation}
Using the expression for $h$ from Eq. (\ref{hubbleina}) for
$\zeta=\zeta_0$, these parameters become,
\begin{equation}
\begin{split}
r=\frac{3(\tilde{\zeta}_0-3)}{4h^2}
a^{-\frac{3}{2}}[\frac{\tilde{\zeta}_0}{3}-2h]+
\frac{3(\tilde{\zeta}_0-3)}{2h}a^{-\frac{3}{2}}+1
\end{split}
\end{equation}
\begin{equation}
s=-\frac{\frac{3(\tilde{\zeta}_0-3)}{4h^2}
a^{-\frac{3}{2}}[\frac{\tilde{\zeta}_0}{3}-2h]+
\frac{3(\tilde{\zeta}_0-3)}{2h}a^{-\frac{3}{2}}}{
\frac{3(\tilde{\zeta}_0-3)}{2h}a^{-\frac{3}{2}}+\frac{9}{2}}.
\end{equation}
 The $\{r,s\}$ plane
trajectory of the model with $\zeta=\zeta_0$ is shown in figure
\ref{fig:statefinder1}.

\begin{figure}[tbp]
\centering
\includegraphics[width=.45\textwidth]{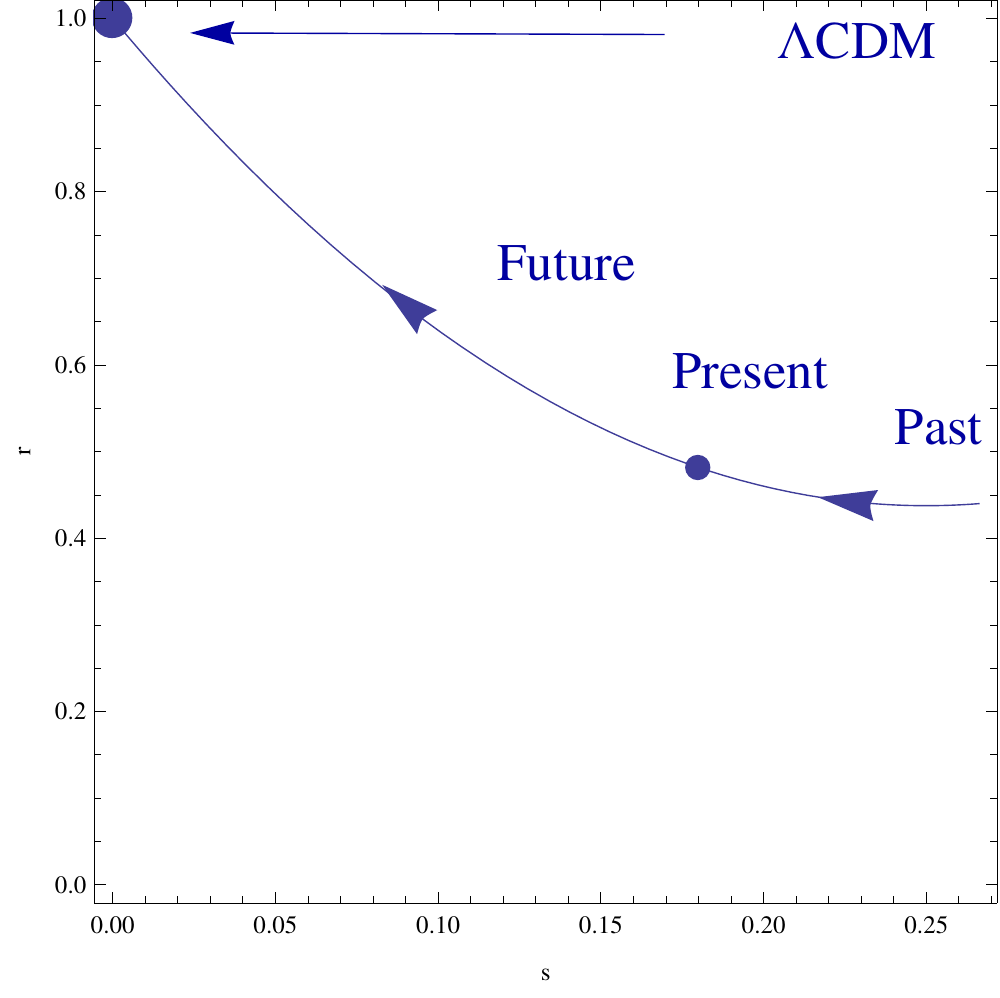}
\caption{\label{fig:statefinder1} The evolution of the model in the
r-s plane for the best estimates of the parameter $\tilde{\zeta}_0$}
\end{figure}

The plot lie in the region $r<1, s>0$, which is the general behavior
of any quintessence model.

\end{enumerate}


\begin{thebibliography}{99}

\bibitem{Riess1}
A. G. Riess et al., \emph{Observational Evidence from Supernovae for
an Accelerating Universe and a Cosmological Constant}, \emph{Astron.
J.} {\bf 116} (1998) 1009.

\bibitem{Perl1}
S. Perlmutter et al., \emph{Measurements of $\Omega$ and $\Lambda$
from 42 High-Redshift Supernovae}, \emph{Astrophys. J.} {\bf 517}
(1999) 565.

\bibitem{Bennet1}
C. L. Bennett et al., \emph{First-Year Wilkinson Microwave
Anisotropy Probe (WMAP) Observations: Preliminary Maps and Basic
Results}, \emph{Astrophys. J. Suppl. Ser.} {\bf 148} (2003) 1.

\bibitem{Tegmark1}
Tegmark et al., \emph{Cosmological parameters from SDSS and WMAP},
\emph{Phys. Rev. D} {\bf 69} (2004) 103501.

\bibitem{Seljak}
Seljak et al., \emph{Cosmological parameter analysis including SDSS
Ly$\ensuremath{\alpha}$ forest and galaxy bias: Constraints on the
primordial spectrum of fluctuations, neutrino mass, and dark
energy}, \emph{Phys. Rev. D} {\bf 71} (2005) 103515.

\bibitem{Komatsu1}
E. Komatsu et al., \emph{Seven-year Wilkinson Microwave Anisotropy
Probe (WMAP) Observations: Cosmological Interpretation},
\emph{Astrophys. J. Suppl. Ser.} {\bf 192} (2011) 18.

\bibitem{Weinberg1}
Steven Weinberg, \emph{The cosmological constant problem},
\emph{Rev. Mod. Phys.} {\bf 61} (1989) 1.

\bibitem{Copeland1}
 Edmund J Copeland, M. Sami and S. Tsujikawa, \emph{Dynamics of
dark energy}, \emph{Int. J. Mod Phys D} {\bf 15} (2006) 1753.

\bibitem{fujii}
Yasunori Fujii, \emph{Origin of the gravitational constant and
particle masses in a scale-invariant scalar-tensor theory},
\emph{Phys. Rev. D} {\bf 26} (1982) 2580.

\bibitem{carroll}
Sean M. Carroll, \emph{Quintessence and the Rest of the World:
Suppressing Long-Range Interactions}, \emph{Phys. Rev. Lett.} {\bf
81} (1998) 3067.

\bibitem{ford}
L. H. Ford, \emph{Cosmological-constant damping by unstable scalar
fields}, \emph{Phys. Rev. D} {\bf 35} (1987) 2339.

\bibitem{Liddle}
Edmund J. Copeland and Andrew R. Liddle and David Wands,
\emph{Exponential potentials and cosmological scaling solutions},
\emph{Phys. Rev. D} {\bf 57} (1998) 4686.

\bibitem{chiba1}
Takeshi Chiba, Takahiro Okabe and Masahide Yamaguchi,
\emph{Kinetically driven quintessence}, \emph{Phys. Rev. D} {\bf 62}
(2000) 023511.

\bibitem{Picon2000}
C. Armendariz-Picon, V. Mukhanov and Paul J. Steinhardt,
\emph{Dynamical Solution to the Problem of a Small Cosmological
Constant and Late-Time Cosmic Acceleration}, \emph{Phys. Rev. Lett.}
{\bf 85} (2000) 4438.

\bibitem{kamen1}
Alexander Kamenshchik, Ugo Moschella and Vincent Pasquier, \emph{An
alternative to quintessence}, \emph{Phys. Lett. B} {\bf 511} (2001)
265.

\bibitem{Bento2002}
M. C. Bento, O. Bertolami and A. A. Sen, \emph{Generalized Chaplygin
gas, accelerated expansion, and dark-energy-matter unification},
\emph{Phys. Rev. D} {\bf 66} (2002) 043507.

\bibitem{capo1}
Salvatore Capozziello, \emph{Curvature quintessence}, \emph{Int. J.
Mod Phys D} {\bf 11} (2002) 483.

\bibitem{Thomas2010}
Thomas P. Sotiriou and Valerio Faraoni, \emph{$f(R)$ theories of
gravity}, \emph{Rev. Mod. Phys.} {\bf 82} (2010) 451.

\bibitem{ferraro1}
R. Ferraro and F. Fiorini, \emph{Modified teleparallel gravity:
Inflation without an inflaton}, \emph{Phys. Rev. D} {\bf 75} (2007)
084031.

\bibitem{Ratbay2011}
Ratbay Myrzakulov, \emph{Accelerating universe from F(T) gravity},
\emph{Eur. Phys. J. C} {\bf 71} (2011) 1752.

\bibitem{nojiri}
Shin'ichi Nojiri, Sergei D. Odintsov, and Misao Sasaki,
\emph{Gauss-Bonnet dark energy}, \emph{Phys. Rev. D} {\bf 71} (2005)
123509.

\bibitem{pad2}
T. Padmanabhan and D. Kothawala, \emph{Lanczos-Lovelock models of
gravity}, \emph{Phys. Rep.} {\bf 531} (2013) 115.

\bibitem{horava1}
Petr Ho\ifmmode \check{r}\else \v{r}\fi{}ava, \emph{Quantum gravity
at a Lifshitz point}, \emph{Phys. Rev. D} {\bf 79} (2009) 084008.

\bibitem{amendola1}
Luca Amendola, \emph{Scaling solutions in general nonminimal
coupling theories}, \emph{Phys. Rev. D} {\bf 60} (1999) 043501.

\bibitem{dvali1}
Gia Dvali, Gregory Gabadadze and Massimo Porrati, \emph{4D gravity
on a brane in 5D Minkowski space}, \emph{Phys. Lett. B} {\bf 485}
(2000) 208.

\bibitem{pad}
T. Padmanabhan and SM Chitre, \emph{Viscous universes}, \emph{Phys.
Lett. A} {\bf 120} (1987) 433.

\bibitem{waga}
Ioav Waga and Rossana C. Falc\~ao and Rajat Chanda,
\emph{Bulk-viscosity-driven inflationary model}, \emph{Phys. Rev. D}
{\bf 33} (1986) 1839.

\bibitem{Cheng}
Baolian Cheng, \emph{Bulk viscosity in the early universe},
\emph{Phys. Lett. A} {\bf 160} (1991) 329.

\bibitem{fabris1}
J. C. Fabris, and S. V. B. Gonçalves, and R.de Sá Ribeiro,
\emph{Bulk viscosity driving the acceleration of the Universe},
\emph{Gen. Relat. Gravit.} {\bf 38} (2006) 495.

\bibitem{li1}
Baojiu Li and John D. Barrow, \emph{Does bulk viscosity create a
viable unified dark matter model?}, \emph{Phys. Rev. D} {\bf 79}
(2009) 103521.

\bibitem{Hiplito1}
W. S. Hip\'olito-Ricaldi and H. E. S. Velten, and  W. Zimdahl,
\emph{Viscous dark fluid universe}, \emph{Phys. Rev. D} {\bf 82}
(2010) 063507.

\bibitem{av1}
Arturo Avelino and Ulises Nucamendi, \emph{Can a matter-dominated
model with constant bulk viscosity drive the accelerated expansion
of the universe?}, \emph{JCAP} {\bf 04} (2009) 006.

\bibitem{av2}
Arturo Avelino and Ulises Nucamendi, \emph{Exploring a
matter-dominated model with bulk viscosity to drive the accelerated
expansion of the Universe}, \emph{JCAP} {\bf 08} (2010) 009.

\bibitem{wilson1}
James R. Wilson, Grant J. Mathews, and George M. Fuller, \emph{Bulk
viscosity, decaying dark matter, and the cosmic acceleration},
\emph{Phys. Rev. D} {\bf 75} (2007) 043521.

\bibitem{kowalski1}
M. Kowalski et al., \emph{Improved Cosmological Constraints from
New, Old, and Combined Supernova Data Sets}, \emph{Astrophys. J.}
{\bf 686} (2008) 749.

\bibitem{athira}
Athira Sasidharan and Titus K. Mathew, \emph{Bulk viscous matter and
recent acceleration of the universe}, \emph{Eur. Phys. J. C} {\bf
75} (2015) 348.

\bibitem{WMAP2009}
G. Hinshaw et. al., \emph{Five-Year Wilkinson Microwave Anisotropy
Probe Observations: Data Processing, Sky Maps, and Basic Results},
\emph{Astrophys. J. Suppl. S} {\bf 180} (2009) 225.

\bibitem{Eckart1}
Carl Eckart, \emph{The Thermodynamics of Irreversible Processes.
III. Relativistic Theory of the Simple Fluid}, \emph{Phys. Rev.}
{\bf 58} (1940) 919.

\bibitem{Israel2}
W. Israel and J. M. Stewart, \emph{Transient relativistic
thermodynamics and kinetic theory}, \emph{Ann. Phys.} {\bf 118}
(1979) 341.

\bibitem{Israel3}
W. Israel and J. M. Stewart, \emph{On transient relativistic
thermodynamics and kinetic theory. II}, \emph{Proc. R. Soc. Lond. A
Math. Phys. Sci.} {\bf 365} (1979) 43.

\bibitem{kremer2003}
G. M. Kremer and F. P. Devecchi, \emph{Viscous cosmological models
and accelerated universes}, \emph{Phys. Rev. D} {\bf 67} (2003)
047301.

\bibitem{titus}
Titus K Mathew, M B Aswathy, M Manoj, \emph{Cosmology and
thermodynamics of FLRW universe with bulk viscous stiff fluid},
\emph{Eur. Phys. J. C} {\bf 74} (2014) 3188.

\bibitem{Hu2006}
Ming-Guang Hu and Xin-He Meng, \emph{Bulk viscous cosmology:
statefinder and entropy}, \emph{Phys. Lett. B} {\bf 635} (2006) 186.

\bibitem{Gron2013}
N Mostafapoor and O Gron, \emph{Viscous $\Lambda$CDM Universe
Models}, \emph{Astrophys. Space Sci.} {\bf 333} (2011) 357.

\bibitem{Brevik2005}
I. Brevik and O. Gorbunova, \emph{Dark energy and viscous
cosmology}, \emph{Gen. Rel. Grav.} {\bf 37} (2005) 2039.

\bibitem{ren1}
Jie Ren and Xin-He Meng, \emph{Cosmological model with viscosity
media (dark fluid) described by an effective equation of state},
\emph{Phys. Lett. B} {\bf 633} (2006) 1.

\bibitem{Chimento1997}
Luis P. Chimento and Alejandro S. Jakubi, \emph{Dissipative
cosmological solutions}, \emph{Class. Quantum Grav.} {\bf 14} (1997)
7.

\bibitem{Colistete2007}
R. Colistete, Jr., J. C. Fabris, J. Tossa, and W. Zimdahl,
\emph{Bulk viscous cosmology}, \emph{Phys. Rev. D} {\bf 76} (2007)
103516.

\bibitem{Lie} L. N. Liebermann, \emph{The second viscosity of liquids}, \emph{Phys. Rev.} {\bf 75} 1415 (1949)

\bibitem{Singh} J.P. Singh, Pratibha Singh, Raj Bali,
\emph{ Bulk viscosity and decaying vacuum density in Friedmann
universe}, \emph{Int J Theor Phys}, {\bf 51} 3828 (2012)

\bibitem{Avelino} A. Avelino et.al, \emph{Bulk Viscous Matter-dominated Universes: Asymptotic Properties},
 \emph{JCAP}, {\bf 1308} 12 (2013)

\bibitem{Nojiri} Shinichi Nojiri and Sergei D. Odintsov, \emph{Inhomogeneous Equation of State of the Universe:  Phantom Er
a, Future Singularity and Crossing the Phantom Barrier}, \emph{Phys.
Rev. D} {\bf 72} 023003 (2005)

\bibitem{Nojiri1} S. Capozziello, V. F. Cardone, E. Elizalde, S. Nojiri and S. D. Odintsov,
\emph{Inhomogeneous Equation of State of the Universe:  Phantom Er
a, Future Singularity and Crossing the Phantom Barrier}, \emph{Phys.
Rev.} {\bf 73} 043512 (2006)

\bibitem{Xu} Xu Dou and Xin-He Meng, \emph{Scalar perturbation of the viscosity dark fluid cosmological mode}, \emph{}arXiv:0911.5401

\bibitem{Velton2011}
Hermano Velten, Dominik J. Schwarz, \emph{Constraints on dissipative
unified dark matter}, \emph{JCAP} {\bf 09} (2011) 016.

\bibitem{weinberg2} S. Weinberg, \emph{Gravitation and
cosmology: principles and applications of the general theory of
relativity}, John Wiley \& sons Inc., New york U.S.A. (1972).

\bibitem{Bento} M. C. Bento, O. Bertolami and A.A. Sen, \emph{Generalized Chaplygin Gas, Accelerated Expansion and Dark Energy-Matter
Unification}, \emph{Phys. Rev. D} {\bf 66} (2002) 043507.

\bibitem{Andrea} A. Addazi, A. Marcian and S. Alexander,
\emph{A Unified picture of Dark Matter and Dark Energy from
Invisible QCD}, \emph{arXiv: 1603.01853} (2016).

\bibitem{Gorini} V. Gorini et.al., \emph{The Chaplygin gas as a model for dark energy}, \emph{arXiv:0403062} (2004).
\bibitem{Zimdahl} W. Zimdahl et. al., \emph{Viscous dark fluid unierse: A unified model of the dark sector?},
 \emph{Int. J. Mod. Phys. Conf. Ser.} {\bf 03} (2011) 312.

\bibitem{Brevik} I. Brevik O. Gr{\o}n, \emph{Universe Models with Negative Bulk Viscosity},
\emph{Astrophys. Space Sci.} {\bf 347} 399 (2013)

\bibitem{Radi1} N. Radicella and D. Pav$\acute{o}$n, \emph{The generalized second law in universes
with quantum corrected entropy relations}, \emph{Phys. Lett. B} {\bf
691}, 121 (2010)

\bibitem{Radi2} N.Radicella and D. Pav$\acute{o}$n, \emph{A thermodynamic motivation for dark energy},
\emph{Gen. Rel. Grav.} {\bf 44}, 685 (2012)

\bibitem{Li} En-Kun Li, Yu Zhang, Jin-Ling Geng, and Peng-Fe Duan, \emph{Generalized holographic Ricci dark energy
and generalized second law of thermodynamics in Bianchi Type I
universe} \emph{Gen.Rel.Grav.} {\bf 47} 136 (2015)

\bibitem{karami} K Karami, S Ghaffari and M M Soltanzadeh, \emph{The generalized second law of gravitational
thermodynamics on the apparent and event horizons in FRW cosmology}
\emph{Class. Quantum Grav.} {\bf 27} 205021 (2010)

\bibitem{Bousso} R. Bousso and N. Engelhardt, \emph{Generalized Second Law for
Cosmology} \emph{Phys. Rev. D} {\bf 93} 024025 {2016}

\bibitem{Zhov} Jia Zhou et.al., \emph{The generalized second law of thermodynamics in the
accelerating universe}, \emph{Phys. Lett. B} {\bf 652}, 86 (2007)

\bibitem{Setare} M. R. Setare and A. Sheykhi, \emph{Viscous dark energy and generalized second law of thermodynamics},
\emph{Int. J. Mod. Phys. D} {\bf 19}, 1205 (2010)

\bibitem{Wang} B. Wang, Y. Gong and E Abdalla, \emph{Thermodynamics of an accelerated expanding universe},
\emph{Phys. Rev. D} {\bf 74}, 083520 (2006)

\bibitem{Akbar} M. Akbar, \emph{Viscous Cosmology and Thermodynamics of Apparent Horizon},
\emph{Chin. Phys. Lett.} {\bf 25}, 4199 (2008)

\bibitem{Titus} Titus K. Mathew, P. Praseetha, \emph{Holographic dark energy and generalized second law},
\emph{Mod. Phys. Lett. A} {\bf 29}, 1450023 (2014)

\bibitem{Praseetha} P Praseetha and Titus K Mathew, \emph{Entropy of holographic dark energy and the
generalized second law}, \emph{Class. Quantum Grav.} {\bf 31},
185012 (2014)

\bibitem{Sheykhi1} A. Sheykhi, B. Wang and R. G. Cai, \emph{Thermodynamical properties of apparent horizon in warped DGP braneworld},
\emph{Nucl. Phys. B} {\bf 779}, 1 (2007)

\bibitem{Sheykhi2} A. Sheykhi, B. Wang and R. G. Cai, \emph{Deep connection between thermodynamics and gravity in Gauss-Bonnet braneworlds},
\emph{Phys. Rev. D} {\bf 76}, 023515 (2007)

\bibitem{Sahni} V. Sahni, T.D. Saini, A.A. Starobinsky, U. Alam, \emph{Statefinder-A new geometrical diagnostic of dark
energy}, \emph{JETP Lett.} {\bf 77} 201 (2003)



\end{thebibliography}
\end{document}